\documentclass[aps,pre,showpacs,twocolumn]{revtex4}
\usepackage{bm}
\usepackage{graphicx}
\bibstyle{approve.bib}
\usepackage{amssymb}
\usepackage{amsmath}
\usepackage{esint}
\usepackage{epstopdf}
\usepackage{color}
\usepackage{gensymb}
\usepackage{mathtools}
\usepackage{suffix}
\usepackage{fancyhdr}

\newcommand{\be}{\begin{equation}} 
\newcommand{\ee}{\end{equation}}
\newcommand{\bea}{\begin{eqnarray}}
\newcommand{\eea}{\end{eqnarray}}
\newcommand{\ph}{\mathbf{\Psi}}
\newcommand{\lan}{\left\langle}
\newcommand{\ran}{\right\rangle}
\newcommand{\br}{\mathbf{r}}

\newcommand{\hd}{\hat{n}}

\newcommand{\bq}{\mathbf{q}}

\newcommand{\F}{\mathbf{F}}

\newcommand{\e}{\varepsilon}

\newcommand{\tv}{\tilde{v}}

\newcommand{\may}{\tilde{h}_{ij}(q)}

\newcommand{\tG}{\tilde{G}}

\newcommand{\m}{_{\rm m}}
\newcommand{\ce}{_{\rm c}}
\newcommand{\G}{_{\rm G}}

\newcommand{\h}{_{\rm h}}
\newcommand{\n}{_{\rm n}}

\newcommand{\hn}{\hat{n}}

\begin{document}

\title{Systematic incorporation of the ionic hard-core size into the Debye-H\"{u}ckel theory via the cumulant expansion of the Schwinger-Dyson equations}

\author{Sahin Buyukdagli}
\address{Department of Physics, Bilkent University, Ankara 06800, Turkey}

\begin{abstract}

\noindent{\bf ABSTRACT:}  The Debye-H\"{u}ckel (DH) formalism of bulk electrolytes equivalent to the gaussian-level closure of the electrostatic Schwinger-Dyson (SD) identities without the interionic hard-core (HC) coupling is extended via the cumulant treatment of these equations augmented by HC interactions.  By confronting the monovalent ion activity and pressure predictions of our cumulant-corrected DH (CCDH) theory with hypernetted-chain (HNC) results and Monte-Carlo (MC) simulations from the literature, we show that this rectification extends the accuracy of the DH formalism from submolar into molar salt concentrations.  In the case of internal energies or the general case of divalent electrolytes mainly governed by charge correlations, the improved accuracy of the CCDH theory is limited to submolar ion concentrations. Comparison with experimental data from the literature shows that via the adjustment of the hydrated ion radii, the CCDH formalism can equally reproduce the non-uniform effect of salt increment on the ionic activity coefficients up to molar concentrations. The inequality satisfied by these HC sizes coincides with the cationic branch of the Hofmeister series.
\end{abstract}

\date{\today}
\maketitle   

\section{Introduction}

The repulsive core interactions between the elementary constituents of matter play a central role in the setting of the stability conditions in the Universe. From the electrostatic forces governing ordinary matter at the molecular scale~\cite{Dyson} to the intense gravitational forces dominating astronomical objects~\cite{OppenAstro}, HC interactions counterbalancing attractive forces of various origins prevent packed matter from collapsing in on itself at high densities.

In nanoscale systems  incorporating charged fluids, electrostatic forces attenuated by HC interactions drive various salt-regulated mechanisms essential to the persistance and evolution of life on Earth, such as the transport of ions and viral pathogens through nuclear pores~\cite{Bont,Holm}, and the affinity of DNA molecules to biological membranes as well as their stable wrapping around histone proteins~\cite{gn1}. The collective effect of electrostatic and HC interactions is also relevant to nanofluidic charge transport and energy conversion techniques~\cite{Boc1,Boc2}, the functioning of energy storage devices~\cite{rev2},  the manufacturing of synthetic fuels~\cite{sf}, liquid purification by flocculation~\cite{pur}, and water desalination by nanofiltration~\cite{Yar,Szymczyk}. 

In solutions containing ions with typical hydrated sizes, HC interactions becoming substantial at submolar salt concentrations largely dominate the thermodynamics of the liquid in the molar concentration regime~\cite{Book}. This hierarchy limits the validity of the predictions by purely electrostatic theories such as the original DH formalism~\cite{DH} to dilute salt concentrations. Hence, the characterization of the aforementioned nanoscale systems often involving charged solutes at molar concentrations necessitates the incorporation of the HC and electrostatic interactions into electrolyte models on an equal footing. 

In this regard, MC simulation techniques have been particularly reliable tools for accessing the molar concentration regime, thereby enabling the verification of the electrolyte models via direct confrontation with experimental data~\cite{Valleau,Svensson,NetzMC,NetzMC2,LevinMC}. In addition, classical density functional theories and integral equation formalisms with substantially lower computational cost have proved to be accurate alternatives to MC simulations~\cite{Hansen}.  

The explicit solution of the Ornstein-Zernike (OZ) equation belonging to the second category requires a closure relation for the pair correlation function. The first technical issue encountered at this level is the lack of a systematic and controlled method for the choice of this approximate closure. Moreover, in the case of charged liquids governed by long-range interactions, closure relations beyond the weak-coupling (WC) random phase and mean-spherical approximations~\cite{Blum,Henderson,Boda} lead to involved solution schemes that can shadow the analytical transparency of the underlying physics~\cite{Hoye,Hansen,Roji}. This indicates the necessity of complementing the integral equation theories by simpler theoretical methods providing analytical insight into the macroscopic effects originating from the intermolecular interactions in the solution.

Along these lines, Attard~\cite{AttardPRE,AttardJCP} and Kjellander~\cite{Kj1,Kj2,KjJCP2016,Kj2020} have developed ingenious theoretical approaches enabling the analytically transparent characterization of the liquid thermodynamics at solute concentrations inaccessible by the DH formalism. Namely, by applying the global electroneutrality constraint and the second moment condition~\cite{secmom} to generalized DH-like charge distribution functions, they derived effective screening parameters, and evaluated the thermodynamic functions of bulk liquids in the large density regime governed by the close competition between electrostatic and HC correlations.  The accuracy of these generalized DH formalisms has been confirmed via extensive comparison of their predictions with MC simulations and the quasi-exact numerical solution of the OZ equation with HNC closures.  In particular, Kjellander's modified DH theory has been shown to reproduce accurately the Kirkwood transition~\cite{Kirk1,Kirk2} from the exponentially decaying to the damped oscillatory regime of the charge distribution function~\cite{Kj2020,NetzMC,AttardPRE,AttardJCP}.

In the context of conventional statistical mechanics, the field theory approach to electrostatic interactions has emerged as an efficient and convenient alternative to the aforementioned calculation techniques~\cite{PodWKB}. Owing to its formulation as an explicit partition function, the technical advantages provided by the field theory framework are numerous. First of all, this approach allows the treatment of many-body interactions via systematic and controlled perturbation techniques~\cite{NetzLoop,NetzSC,NetzVir,Podgornik2010} as well as self-consistent computation schemes~\cite{netzvar,HatloVar,Buyuk2020}. Then, the field theory formalism enables the straightforward incorporation of the specific molecular details of electrolytes beyond the primitive model, such as structured solute charges~\cite{Demery,Buyuk2023}, explicit solvent molecules~\cite{DuncSolv,NL1}, structured membranes~\cite{dipmem}, and polyelectrolytes with conformational degrees of freedom~\cite{RudiPol,NetzPol}.

Due to the onset of the competition between the electrostatic and HC interactions at submolar salt concentrations~\cite{Book}, the validity of the field-theoretic models neglecting the {\it pairwise HC interactions} is limited to dilute salt solutions. In order to overcome this limitation, in this work, we formulate a field-theoretic calculation scheme that enables the evaluation of thermodynamic averages by treating the Coulombic and HC interactions between all ion species on an equal footing. The corresponding CCDH formalism is based on the {\it cumulant expansion} of the liquid partition function together with the SD identities~\cite{justin}. In our earlier works, the {\it virial expansion} of these identities~\cite{Buyuk2020} whose validity requires the dilute salt condition has been used  to characterize correlation effects in nanofluidic ion transport~\cite{Heyden2005,JPCB2020}, polymer translocation~\cite{BuyukLang2022}, and salt-induced dielectric decrement in polar liquids~\cite{BuyukDielDec}. The cumulant expansion method adopted in the present work allows us to avoid precisely the dilute salt constraint accompanying the virial approximation used in the aforementioned studies.

By confronting the CCDH formalism with MC and HNC data from the literature~\cite{NetzMC,Svensson,Valleau,AttardPRE}, we show that the incorporation of the HC ion size extends the accuracy of the DHLLs for the osmotic pressure and the ionic activity coefficients of monovalent salt solutions from submolar into molar salt concentrations. This substantial upgrade is the main progress of our work. Throughout the article, we also identify systematically the limitations of our cumulant calculation scheme, and elaborate potential improvements and applications in Conclusions.

\section{Theory}

We introduce here our electrolyte model, and derive its partition function in the form of a functional integral over fluctuating electrostatic and HC potentials. Then, by exploiting the invariance of this partition function with respect to the variation of the electrostatic potential, we derive the electrostatic SD equations. By solving the resulting identities via a first order cumulant expansion, we obtain the electrostatic Green's function, and calculate the two point distribution functions characterizing ion correlations. In Sec.~\ref{epr}, these distribution functions will be used for the computation of the excess energy, the pressure, and the ionic activity coefficients.

\subsection{Electrolyte model and liquid partition function}
\label{mod}

The electrolyte model is composed of $p$ ion species located in an implicit solvent of uniform permittivity $\e_{\rm w}$. The ions of the species $i$ with fugacity $\lambda_i$ and concentration $ n_i$ are point charges of valency $q_i$ placed at the center of a HC sphere with diameter $d$. Thus, the interactions between two ions separated by the distance $r$ are set by the HC potential $v\h(r)$ defined for all species as
\be\label{eq5}
e^{-v\h(\br)}=\theta(r-d),
\ee
where $\theta(x)$ is the Heaviside step function~\cite{math}, and the bulk Coulomb potential $v\ce(\br)=\ell_{\rm B}/r$ characterized by the Bjerrum length $\ell_{\rm B}=e^2/(4\pi\e_{\rm w}k_{\rm B}T)$, where $e$ stands for the electron charge, $k_B$ is the Boltzmann constant, and $T$ is the liquid temperature.

We derive now the partition function of this electrolyte in a functional integral form. The grand canonical (GC) partition function of the liquid is given by the trace of the Boltzmann distribution function over the particle positions and number fluctuations,
\bea
\label{eq1}
Z\G=\prod_{i=1}^p\sum_{N_i=1}^\infty\frac{\lambda_i^{N_i}}{N_i!}\prod_{j=1}^p\prod_{k=1}^{N_j}\int\mathrm{d}^3\br_{jk}e^{-\beta(E\m+E\n)}.
\eea
In Eq.~(\ref{eq1}), the pairwise ionic interaction energy is
\bea
\label{eq2}
\beta E\m&=&\frac{1}{2}\sum_{i=1}^p\sum_{j=1}^p\sum_{k=1}^{N_i}\sum_{l=1}^{N_j}\left[q_iq_jv\ce(\br_{ik}-\br_{jl})\right.\\
&&\hspace{2.65cm}\left.+v\h(\br_{ik}-\br_{jl})\right].\nonumber
\eea
Moreover, the single-body energy component reads
\be
\label{eq3}
\beta E\n=\sum_{i=1}^p\sum_{j=1}^{N_i}w_i(\br_{ij})-\sum_{i=1}^pN_i\epsilon_i,
\ee
where the external potential $w_i(\br)$ will enable us to derive the density distributions, and the self-energy component
\be
\epsilon_i=\frac{1}{2}\left[q_i^2v\ce(0)+v\h(0)\right]
\ee
has been subtracted from the Hamiltonian. 

Using now the charge and number density operators
\be
\label{eq6}
\hn\ce(\br)=\sum_{i=1}^p\sum_{j=1}^{N_i}q_i\delta^3(\br-\br_{ij});\;\hn\h(\br)=\sum_{i=1}^p\sum_{j=1}^{N_i}\delta^3(\br-\br_{ij}),
\ee
the pairwise interaction energy~(\ref{eq2}) can be recast as
\bea
\label{eq8}
\beta E\m&=&\frac{1}{2}\int\mathrm{d^3}\br\mathrm{d^3}\br'\left[\hn\ce(\br)v\ce(\br-\br')\hn\ce(\br')\right.\\
&&\left.\hspace{2cm}+\hn\h(\br)v\h(\br-\br')\hn\h(\br')\right]\nonumber.
\eea
Next, we switch from the particle density to the field representation via an Hubbard-Stratonovich (HS) transformation for each type of pairwise interaction,
\bea\label{eq9}
\hspace{-5mm}&&e^{-\frac{1}{2}\int\mathrm{d}^3\br\mathrm{d}^3\br'\hn_\gamma(\br)v_\gamma(\br-\br')\hn_\gamma(\br')}\\
\hspace{-5mm}&&=\int\mathcal{D}\psi_\gamma\;e^{-\frac{1}{2}\int\mathrm{d}^3\br\mathrm{d}^3\br'\psi_\gamma(\br)v^{-1}_\gamma(\br-\br')\psi_\gamma(\br')}e^{i\int\mathrm{d}^3\br c_\gamma(\br)\psi_\gamma(\br)},\nonumber
\eea
where $\psi_\gamma(\br)$ stand for the fluctuating potentials associated with the Coulomb ($\gamma={\rm c}$) or HC coupling ($\gamma={\rm h}$). This transformation enables us to evaluate exactly the geometric sum in Eq.~(\ref{eq1}), and to recast the GC partition function as a double functional integral of the form
\be
\label{eq12}
Z\G=\int\mathcal{D}\ph\;e^{-\beta H[\ph]},
\ee
where we used the shorthand vector notations for the fluctuating potentials $\ph=(\psi\ce,\psi\h)$ and the functional integration measure $\mathcal{D}\ph=\mathcal{D}\psi\ce\mathcal{D}\psi\h$.

In Eq.~(\ref{eq12}), the Hamiltonian functional is defined as
\bea
\label{eq13}
\beta H[\ph]&=&\frac{1}{2}\sum_{\gamma={\rm c,h}}\int\mathrm{d}^3\br\mathrm{d}^3\br'\psi_\gamma(\br)v^{-1}_\gamma(\br-\br')\psi_\gamma(\br')\nonumber\\
&&-\sum_{i=1}^p\int\mathrm{d}^3\br\;\hd_i(\br).
\eea
The first term on the r.h.s. of Eq.~(\ref{eq13}) incorporates the energy of the quadratic potential fluctuations. Then, the second line of Eq.~(\ref{eq13})  includes the fluctuating ion density
\be
\label{eq14}
\hd_i(\br)=\lambda_i e^{\epsilon_i-w_i(\br)}e^{i\psi\h(\br)+iq_i\psi\ce(\br)}.
\ee
Unless stated otherwise, in the remainder, we will set $w_i(\br)=0$.

\subsection{Derivation of the electrostatic SD identities}
\label{ap1}

We review here the derivation of the SD equations~\cite{justin} introduced for charged systems in Ref.~\cite{Buyuk2020}. To this aim, we define first the functional integral
\be\label{eq15}
I=\int\mathcal{D}\ph\;e^{-\beta H[\ph]}F[\ph]
\ee 
including the general functional $F[\ph]$. Next, we introduce an infinitesimal shift of the electrostatic potential, $\psi\ce(\br)\to\psi\ce(\br)+\delta\psi\ce(\br)$, and linearize Eq.~(\ref{eq15}) in terms of the infinitesimally small function $\delta\psi\ce(\br)$. The variation of the integral follows in the form
\bea\label{eq16}
\delta I&=&\int\mathrm{d}\br\delta\psi\ce(\br)\int\mathcal{D}\ph e^{-\beta H[\ph]}\\
&&\hspace{2.3cm}\times\left\{\frac{\delta F[\ph]}{\delta\psi\ce(\br)}-F[\ph]\frac{\delta H[\ph]}{\delta\psi\ce(\br)}\right\}.\nonumber
\eea
At this point, we exploit the invariance of the integral~(\ref{eq15}) under the potential shift $\delta\psi\ce(\br)$. Thus, setting Eq.~(\ref{eq16}) to zero, i.e. $\delta I=0$, and dividing the result by the partition function~(\ref{eq12}), one obtains
\be
\label{eq17}
\lan\frac{\delta F[\ph]}{\delta\psi\ce(\br)}\ran=\lan F[\ph]\frac{\delta H[\ph]}{\delta\psi\ce(\br)}\ran,
\ee
where the field average of the general functional $F[\ph]$ is
\be\label{eq18}
\lan F[\ph]\ran=\frac{1}{Z_{\rm G}}\int\mathcal{D}\ph\;e^{-\beta H[\ph]}F[\ph].
\ee
Finally, specifying the form of the functional in Eq.~(\ref{eq17}) as $F[\ph]=1$ and $F[\ph]=\psi\ce(\br')$, the formally exact SD equations respectively follow as
\be\label{eq19}
\lan\frac{\delta\left(\beta H[\ph]\right)}{\delta\psi\ce(\br)}\ran=0;\hspace{5mm}\lan\frac{\delta\left(\beta H[\ph]\right)}{\delta\psi\ce(\br)}\psi\ce(\br')\ran=\delta^3(\br-\br').
\ee
In order to simplify the notation, from now on, the argument of the functionals will be omitted.

\subsection{Cumulant expansion scheme}
\label{sc}

Due to the non-linear form of the Hamiltonian functional~(\ref{eq13}), the statistical averages defined by Eq.~(\ref{eq18}) cannot be calculated analytically. Thus, we will evaluate these averages within a cumulant expansion scheme. This cumulant approximation is based on the exact splitting of the Hamiltonian into two components as
\be
\label{eq21}
H=H_0+t\delta H,
\ee
with the gaussian component
\bea
\label{eq22}
\beta H_0&=&\int\frac{\mathrm{d}^3\br\mathrm{d}^3\br'}{2}\left[\psi\ce(\br)G^{-1}(\br-\br')\psi\ce(\br')\right.\\
&&\left.\hspace{1.75cm}+\psi\h(\br)v^{-1}\h(\br-\br')\psi\h(\br')\right],\nonumber
\eea
and the non-linear part to be treated perturbatively,
\bea
\label{eq22II}
\beta\delta H&=&\int\frac{\mathrm{d}^3\br\mathrm{d}^3\br'}{2}\psi\ce(\br)\left[v\ce^{-1}-G^{-1}\right]_{\br,\br'}\psi\ce(\br')\nonumber\\
&&-\sum_{i=1}^p\int\mathrm{d}^3\br\;\hd_i(\br).
\eea
In Eq.~(\ref{eq21}), the parameter $t$ to be set to unity will allow to keep track of the cumulant expansion order. Moreover, the unknown electrostatic kernel $G(\br-\br')$ will be determined from the solution of the SD identities in Eq.~(\ref{eq19}).

The main approximations underlying this cumulant approach are (i) the choice of the reference Hamiltonian~(\ref{eq22}) as a purely gaussian functional, which assumes the higher order non-linear fluctuations included in the perturbative correction~(\ref{eq22II}) to be weak, and (ii) the inclusion of the bare HC kernel $v^{-1}_{\rm h}(\br)$ in Eq.~(\ref{eq22}). In the remainder, the consequences and the validity limits of these approximations will be  identified via comparison with MC simulations and HNC results.

Following this cumulant expansion scheme, we substitute now the decomposition~(\ref{eq21}) into Eq.~(\ref{eq18}), and Taylor-expand the result at the order $O(t)$. The statistical average of the general functional $F$ reduces to
\be
\label{eq23}
\lan F\ran=\lan F\ran_0-t\left[\lan\beta\delta H F\ran_0-\lan\beta\delta H\ran_0\lan F\ran_0\right]+O\left(t^2\right),
\ee
where the gaussian-level average is
\be\label{eq24}
\lan F\ran_0=\frac{1}{Z_0}\int\mathcal{D}\ph\;e^{-\beta H_0}F,
\ee
with the partition function $Z_0=\int\mathcal{D}\ph\;e^{-\beta H_0}$. Owing to the gaussian form of the Hamiltonian $H_0$ included in Eq.~(\ref{eq24}), the statistical averages in Eq.~(\ref{eq23}) can be analytically evaluated via the inverse HS transformation
\bea
\label{eq20II}
&&\lan e^{i\int\mathrm{d}\br\left[J\ce(\br)\psi\ce(\br)+J\h(\br)\psi\h(\br)\right]}\ran_0\\
&&=e^{-\frac{1}{2}\int\mathrm{d}^3\br\mathrm{d}^3\br'\left[J\ce(\br)G(\br-\br')J\ce(\br')+J\h(\br)v\h(\br-\br')J\h(\br')\right]}\nonumber
\eea
including the generating functions $J_{\rm c,h}(\br)$.

To recapitulate, the generality of the CCDH scheme with respect to the DH theory is due to the incorporation of the HC interactions in the reference Hamiltonian~(\ref{eq22}), and the inclusion of the cumulant correction~(\ref{eq22II}). Thus, the DHLLs of the thermodynamic functions calculated in the remainder will be reached via the limits of vanishing cumulant correction $t\to0$ and HC size $d\to0$.

\subsection{Electroneutrality condition}

For the evaluation of the first SD identity in Eq.~(\ref{eq19}), we carry out first the formal splitting of the electrostatic kernel into its gaussian and cumulant components,
\be
\label{eq27}
G(\br-\br')=G_0(\br-\br')+t\;G_1(\br-\br')+O\left(t^2\right).
\ee
Then, we evaluate the cumulant expansion~(\ref{eq23}) of the ion density $ n_i(\br)=-\delta Z\G/\delta w_i(\br)=\lan\hd_i(\br)\ran$, i.e.
\be\label{eq27II}
 n_i(\br)\approx\lan \hd_i(\br)\ran_0-t\left[\lan\beta\delta H \hd_i(\br)\ran_0-\lan\beta\delta H\ran_0\lan \hd_i(\br)\ran_0\right].
\ee
Injecting into Eq.~(\ref{eq27II}) the density functional~(\ref{eq14}) and the Hamiltonian component~(\ref{eq22II}), calculating the gaussian averages with Eq.~(\ref{eq20II}), and accounting for the expansion~(\ref{eq27}), one obtains at the cumulant order $O\left(t\right)$
\bea
\label{eq28}
 n_i&\approx&\Lambda_i-t\Lambda_i\left\{\frac{q_i^2}{2}G_1(0)+2\sum_{j=1}^p\Lambda_jB_{ij}\right\}\\
&&-t\Lambda_iq_i^2\int\frac{\mathrm{d}^3\br'\mathrm{d}^3\br''}{2}\left[G_0^{-1}(\br'-\br'')-v\ce^{-1}(\br'-\br'')\right]\nonumber\\
&&\hspace{2.9cm}\times G_0(\br-\br')G_0(\br-\br''),\nonumber
\eea
with the rescaled fugacity and the self-energy defined as
\bea
\label{eq29}
\Lambda_i&=&\lambda_i\;e^{-\frac{q_i^2}{2}\delta G_0(0)};\\
\label{eq29II}
\delta G_0(0)&=&\lim_{\br\to\br'}[G_0(\br-\br')-v\ce(\br-\br')].
\eea
In Eq.~(\ref{eq28}), we also used the virial coefficient $B_{ij}$ including the Mayer function $h_{ij}(\br)$,
\be
\label{eq30}
B_{ij}=-\frac{1}{2}\int\mathrm{d}^3\br\; h_{ij}(\br);\hspace{3mm}h_{ij}(\br)=e^{-v\h(\br)-q_iq_jG_0(\br)}-1.
\ee
Finally, plugging Eqs.~(\ref{eq13}) and~(\ref{eq22II}) into the cumulant expansion~(\ref{eq23}) of the first SD identity in Eq.~(\ref{eq19}), inserting the kernel expansion~(\ref{eq27}), and using Eq.~(\ref{eq28}), the global electroneutrality condition consistently follows as
\be
\label{eq33}
\sum_{i=1}^pq_i n_i=0.
\ee

\subsection{Derivation of the electrostatic Green's function}

In order to determine the Green's function~(\ref{eq27}), we evaluate here the second SD identity in Eq.~(\ref{eq19}). Carrying out the cumulant expansion~(\ref{eq23}) of the latter, injecting into the result the Hamiltonian functionals~(\ref{eq13}) and~(\ref{eq22II}), evaluating the gaussian averages, and accounting for the electroneutrality condition~(\ref{eq33}), after very long algebra, one obtains the Green's equation 
\bea
\label{eq36II}
\hspace{-4mm}&&\int\mathrm{d}^3\br_1v\ce^{-1}(\br,\br_1)G(\br_1,\br')\nonumber\\
\hspace{-4mm}&&+t\int\mathrm{d}^3\br_1\mathrm{d}^3\br_2\mathrm{d}^3\br_3v\ce^{-1}(\br,\br_1)G(\br_1,\br_2)G(\br_3,\br')\nonumber\\
\hspace{-4mm}&&\hspace{2.6cm}\times\left[G^{-1}(\br_2,\br_3)-v\ce^{-1}(\br_2,\br_3)\right]\nonumber\\
\hspace{-4mm}&&+\sum_{i=1}^p\lambda_iq_i^2e^{-\frac{q_i^2}{2}[G(0)-v\ce(0)]}\nonumber\\
\hspace{-4mm}&&\hspace{2mm}\times\left\{G(\br,\br')-t\int\mathrm{d}^3\br_1\mathrm{d}^3\br_2v^{-1}\ce(\br,\br_1)G(\br_1,\br_2)G(\br_2,\br')\right.\nonumber\\
\hspace{-4mm}&&\hspace{7mm}+t\int\mathrm{d}^3\br_1\mathrm{d}^3\br_2\left[G^{-1}(\br_1,\br_2)-v^{-1}\ce(\br_1,\br_2)\right]\nonumber\\
\hspace{-4mm}&&\hspace{9mm}\left.\times\left[G(\br,\br_1)G(\br_2,\br')-\frac{q_i^2}{2}G(\br,\br_1)G(\br,\br_2)G(\br,\br')\right]\right\}\nonumber\\
\hspace{-4mm}&&+t\sum_{i=1}^p\sum_{j=1}^p\lambda_i\lambda_jq_ie^{-\frac{q_i^2+q_j^2}{2}[G(0)-v\ce(0)]}\nonumber\\
\hspace{-4mm}&&\hspace{8mm}\times\int\mathrm{d}^3\br_1\left\{q_iG(\br,\br')\left[e^{-v\h(\br-\br_1)-q_iq_jG(\br-\br_1)}-1\right]\right.\nonumber\\
\hspace{-4mm}&&\hspace{2.3cm}\left.+q_jG(\br_1,\br')e^{-v\h(\br-\br_1)-q_iq_jG(\br-\br_1)}\right\}\nonumber\\
\hspace{-4mm}&&=\delta(\br-\br').
\eea

The conversion of the ion fugacities  in Eq.~(\ref{eq36II}) into concentrations requires the inversion of Eq.~(\ref{eq28}). Plugging into the latter the expansion $\Lambda_i=\Lambda_i^{(0)}+t\Lambda_i^{(1)}+O\left(t^2\right)$, and identifying the expansion terms $\Lambda_i^{(0)}$ and $\Lambda_i^{(1)}$, the ion fugacity follows as a function of concentration as
\bea\label{eq31}
\Lambda_i&=& n_i+t\frac{q_i^2}{2} n_iG_1(0)+2t n_i\sum_{j=1}^p n_jB_{ij}\\
&&+tq_i^2 n_i\int\frac{\mathrm{d}^3\br_1\mathrm{d}^3\br_2}{2}\left[G_0^{-1}(\br_1-\br_2)-v\ce^{-1}(\br_1-\br_2)\right]\nonumber\\
&&\hspace{2.9cm}\times G_0(\br-\br_1)G_0(\br-\br_2).\nonumber
\eea
Finally, inserting the cumulant kernel expansion~(\ref{eq27}) and the fugacity identity~(\ref{eq31}) into Eq.~(\ref{eq36II}), and separating the terms of the cumulant orders $O\left(t^0\right)$ and $O\left(t^1\right)$, one obtains the following differential equations satisfied by the kernel components of different orders,
\bea
\label{eq37}
\hspace{-3mm}&&\int\mathrm{d}^3\br_1v\ce^{-1}(\br-\br_1)G_0(\br_1-\br')+\sum_{i=1}^p n_iq_i^2G_0(\br-\br')\\
\hspace{-3mm}&&=\delta^3(\br-\br');\nonumber\\
\label{eq38}
\hspace{-3mm}&&\int\mathrm{d}^3\br_1v\ce^{-1}(\br-\br_1)G_1(\br_1-\br')+\sum_{i=1}^p n_iq_i^2G_1(\br-\br')\\
\hspace{-3mm}&&=-\sum_{i,j} n_i n_jq_iq_j\int\mathrm{d}^3\br_1\left\{h_{ij}(\br-\br_1)+q_iq_jG_0(\br-\br_1)\right\}\nonumber\\
\hspace{-3mm}&&\hspace{3.5cm}\times G_0(\br_1-\br').\nonumber
\eea

In this work,  the Fourier transform (FT) of the general function $f(\br)$ and its inverse FT are respectively defined as $\tilde{f}(q)=\int\mathrm{d}^3\br\; f(\br)e^{-i\bq\cdot\br}$ and $f(\br)=(2\pi)^{-3}\int\mathrm{d}^3\br\; \tilde{f}(q)e^{i\bq\cdot\br}$. Fourier transforming now Eqs.~(\ref{eq37})-(\ref{eq38}) according to these definitions, the kernel components in Eq.~(\ref{eq27}) follow in the form
\bea
\label{eq39}
\hspace{-6mm}\tG_0(q)&=&\left[\tv^{-1}\ce(q)+\sum_i n_iq_i^2\right]^{-1};\\
\label{eq39II}
\hspace{-6mm}\tG_1(q)&=&-\tG_0^2(q)\sum_{i,j} n_i n_jq_iq_j\left[\may+q_iq_j\tG_0(q)\right].
\eea
In Eqs.~(\ref{eq39})-(\ref{eq39II}), the FT of the Coulomb potential is given by $\tv\ce(q)=\ell_{\rm B}/q^2$, and  the FT of the Mayer function defined in Eq.~(\ref{eq30}) reads
\bea\label{eq43}
\may&=&-\frac{4\pi}{q^3}\left[\sin(qd)-qd\cos(qd)\right]\\
&&+4\pi\int_d^\infty\mathrm{d}rr^2\frac{\sin(qr)}{qr}\left\{e^{-q_iq_jG_0(r)}-1\right\}.\nonumber
\eea
Finally, via the inverse FT of Eq.~(\ref{eq39}), one finds that the Gaussian-level kernel corresponds to the DH potential whose range is set by the Debye screening length $\kappa^{-1}$, 
\be
\label{eq41}
G_0(r)=\frac{\ell_{\rm B}}{r}e^{-\kappa r};\hspace{5mm}\kappa^2=4\pi\ell_{\rm B}\sum_{i=1}^p n_iq_i^2.
\ee

\subsection{Computation of the pair correlation functions}

The evaluation of the thermodynamic functions investigated below requires the knowledge of the pair distribution function between the ions of the species $i$ and $j$,
\be\label{p1}
g_{ij}(\br,\br')=\frac{1}{ n_i n_j}\lan\sum_{k=1}^{N_i}\sum_{l=1}^{N_j}\delta(\br-\br_{ik})\delta(\br'-\br_{jl})\ran_{\rm G},
\ee
where $\lan\cdot\ran_{\rm G}$ denotes the grand-canonical average. Using the identities~(\ref{eq1}) and~(\ref{eq3}), Eq.~(\ref{p1}) can be recast as
\be
\label{p2}
g_{ij}(\br,\br')=-\frac{1}{ n_i n_j}\frac{1}{\beta Z\G}\frac{\delta^2 Z\G}{\delta w_i(\br)\delta w_j(\br')}.
\ee
Plugging now the functional integral form of the partition function~(\ref{eq12}) into Eq.~(\ref{p2}), the latter can be expressed in terms of the field average~(\ref{eq18}) as
\be
\label{p3}
g_{ij}(\br,\br')=\frac{\lambda_i\lambda_j}{ n_i n_j}e^{\epsilon_i+\epsilon_j}\lan e^{i\psi\h(\br)+iq_i\psi\ce(\br)}e^{i\psi\h(\br')+iq_j\psi\ce(\br')}\ran.
\ee

Evaluating now the cumulant expansion~(\ref{eq23}) of Eq.~(\ref{p3}), computing the functional averages with Eqs.~(\ref{eq24})-(\ref{eq20II}), accounting for the cumulant expansion~(\ref{eq27}) of the kernel, and using the fugacity identity~(\ref{eq31}), the pair correlation function defined by the equality $H_{ij}(\br-\br')=g_{ij}(\br-\br')-1$ follows in the form
\bea
\label{p5}
H_{ij}(\br-\br')&=&h_{ij}(\br-\br')\\
&&+t\;T_{ij}(\br-\br')e^{-v\h(\br-\br')-q_iq_jG_0(\br-\br')}.\nonumber
\eea
Eq.~(\ref{p5}) includes the Mayer function introduced in Eq.~(\ref{eq30}), and the auxiliary function
\bea\label{p6}	
\hspace{-4mm}&&T_{ij}(\br-\br')=\sum_{k=1}^p\int\mathrm{d}^3\br_1\left\{h_{ik}(\br-\br_1)h_{kj}(\br_1-\br')\right.\\
\hspace{-4mm}&&\hspace{3.8cm}\left.-q_iq_jq_k^2G_0(\br-\br_1)G_0(\br_1-\br')\right\}\nonumber\\
\hspace{-4mm}&&\hspace{2cm}-q_iq_jG_1(\br-\br').\nonumber
\eea
Finally, via the convolution theorem, and exploiting the spherical symmetry, Eq.~(\ref{p6}) can be expressed solely in terms of the Fourier-transformed functions~(\ref{eq39})-(\ref{eq43}) as
\bea
\label{p7}
T_{ij}(r)&=&\int_0^\infty\frac{\mathrm{d}qq^2}{2\pi^2}\left\{\sum_{k=1}^p\left[\tilde{h}_{ik}(q)\tilde{h}_{kj}(q)-q_iq_jq_k^2\tG_0^2(q)\right]\right.\nonumber\\
&&\left.\hspace{1.8cm}-q_iq_j\tG_1(q)\right\}\frac{\sin(qr)}{qr}.
\eea

\section{Results and discussion}

\label{res}

\begin{figure*}
\includegraphics[width=1.\linewidth]{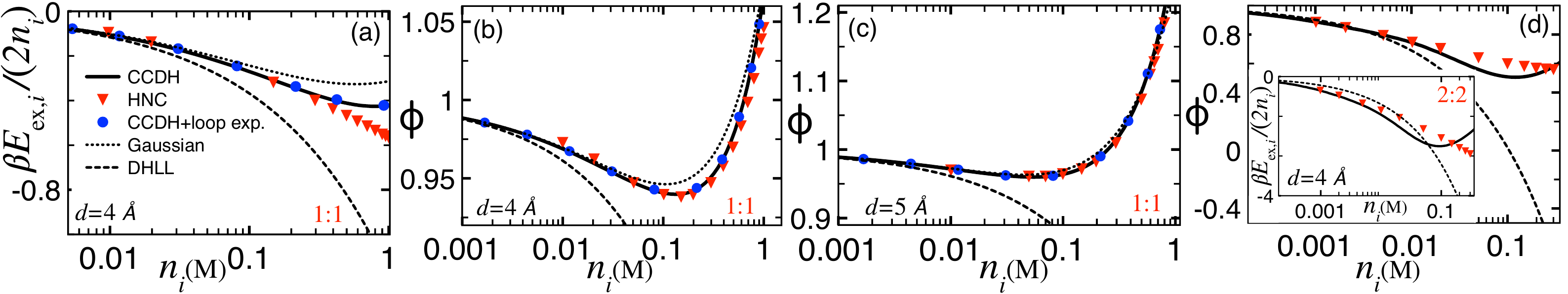}
\caption{(Color online) (a) Energy density and (b)-(c) osmotic coefficient $\phi=\beta P/(2 n_i)$ of 1:1 solutions against salt concentration. (d) Osmotic coefficient (main plot) and excess energy (inset) of 2:2 electrolytes. Solid curves: CCDH predictions from Eqs.~(\ref{p9})-(\ref{p10}) with Eq.~(\ref{p5}).  Red triangles: HNC results from Ref.~\cite{AttardPRE}. Blue disks: closed-form predictions~(\ref{p11})-(\ref{p12}). Dotted curves: gaussian predictions~(\ref{p13})-(\ref{p14}). Dashed curves: DHLL in Eq.~(\ref{p15}). The temperature and the dielectric constant are $T=300$ K and $\e_{\rm w}=78.5$. The HC diameters are displayed in the legends.} 
\label{fig1}
\end{figure*}

Within the framework of the CCDH formalism, we evaluate now the thermodynamic functions of bulk electrolytes and identify the validity regime of our cumulant calculation scheme. To this aim, we confront the predictions of our formalism with the excess energy and osmotic pressure data obtained from MC simulations~\cite{NetzMC,Svensson} and HNC calculations~\cite{AttardPRE}. We also compare our theoretical results with numerical~\cite{NetzMC,Valleau} and experimental ionic activity data~\cite{Exp} from the literature.

\subsection{Excess energy and osmotic coefficient}

\label{epr}

The derivation of the excess energy and pressure for mixed electrolytes is reported in Appendix~\ref{ap0}. In the case of symmetric $q_i:q_i$ solutions, these thermodynamic functions can be expressed in terms of the pair correlation functions in Eq.~(\ref{p5}) as~\cite{Hansen,AttardPRE}
\bea
\label{p9}
\hspace{-4mm}&&\beta E_{\rm ex}=-4\pi\ell_{\rm B}q_i^2 n_i^2\int_d^\infty\mathrm{d}rr\left[H_{+-}(r)-H_{++}(r)\right];\\
\label{p10}
\hspace{-4mm}&&\frac{\beta P}{2 n_i}=1+\frac{\beta E_{\rm ex}}{6 n_i}+\frac{2\pi}{3}d^3 n_i\left[H_{+-}(d^+)+H_{++}(d^+)+2\right].\nonumber\\
\hspace{-4mm}
\eea
Eq.~(\ref{p9}) indicates that the excess energy of the liquid set by the charge correlation function is mainly governed by the electrostatic coupling of opposite charges. Then, according to Eq.~(\ref{p10}), the departure of the pressure from the ideal gas behavior is driven by the competition between these attractive charge correlations (the second term on the r.h.s.), and the repulsive HC correlations embodied by the contact densities (the curly bracket term).

\begin{figure*}
\includegraphics[width=1.\linewidth]{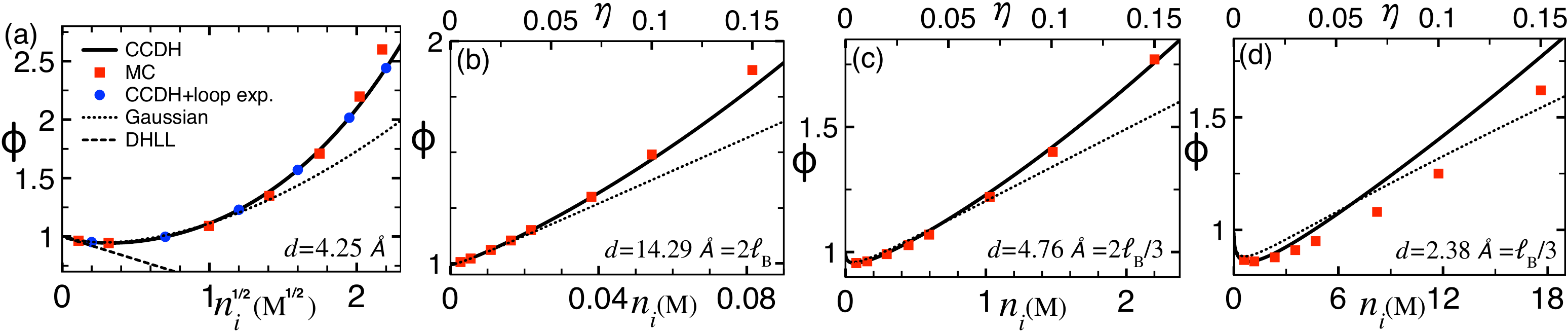}
\caption{(Color online) Osmotic coefficient of 1:1 electrolytes against salt concentration at the HC sizes given in the legends. The symbols and curves have the same signification as in Figs.~\ref{fig1}(b)-(c). The MC data in (a) are from Ref.~\cite{Svensson}. In (b)-(d), the MC data are from Table II~\cite{rem1} of Ref.~\cite{NetzMC}. The upper horizontal axis display the ionic packing fraction $\eta=\pi n_id^3/3$. The temperature and the dielectric constant are $T=298$ K and $\e_{\rm w}=78.5$ in all plots.}  
\label{fig2}
\end{figure*}

In Appendix~\ref{ap3}, via the loop expansion of Eqs.~(\ref{p9})-(\ref{p10}) in terms of the electrostatic coupling parameter 
\be\label{cp}
\Gamma=q_i^2\kappa\ell_{\rm B}, 
\ee
which corresponds to a weak coupling assumption valid for monovalent electrolytes at ambient temperature, we derive the following closed-form equations of state,
\begin{widetext}
\bea\label{p11}
&&\hspace{-1.2cm}\frac{\beta E_{\rm ex}}{2 n_i}=\frac{\beta E_{{\rm ex},0}}{2 n_i}+\frac{t}{96}\left\{6(1+2\kappa d)e^{-2\kappa d}-\left[6+6\kappa d-9\left(\kappa d\right)^2+5\left(\kappa d\right)^3\right]e^{-\kappa d}\right\}\\
&&+\frac{t\Gamma}{32}\left\{(4\kappa d-3)e^{-3\kappa d}+8(2+\kappa d)e^{-2\kappa d}+\left[2\left(\kappa d\right)^2-2\kappa d-13\right]e^{-\kappa d}\right\}+O\left(\Gamma^2\right);\nonumber\\
\label{p12}
\frac{\beta P}{2 n_i}&=&\frac{\beta P_0}{2 n_i}+10t\left(\frac{\pi}{3}n_id^3\right)^2\\
&&+\frac{t}{576}\left\{\left[5\left(\kappa d\right)^4-12\left(\kappa d\right)^2(e^{\kappa d}-2)-12(e^{\kappa d}-1)-2\left(\kappa d\right)^3(5e^{\kappa d}+6)\right]+6\left(\kappa d\right)^2\left(3e^{\kappa d}+1-e^{-2\kappa d}\right)\right\}e^{-2\kappa d}\nonumber\\
&&+\frac{t\Gamma}{96}\left\{(5-4\kappa d)\kappa d-(4\kappa d+3)e^{\kappa d}+\left[16+11\kappa d+6\left(\kappa d\right)^2-2\left(\kappa d\right)^3\right]e^{2\kappa d}+\left[2\left(\kappa d\right)^2-2\kappa d-13\right]e^{3\kappa d}\right\}e^{-4\kappa d}\nonumber\\
&&+O\left(\Gamma^2\right).\nonumber
\eea
\end{widetext}
In Eqs.~(\ref{p11})-(\ref{p12}), the gaussian-level energy and pressure preceding the cumulant corrections of order $O(t)$ read
\bea
\label{p13}
\hspace{-4mm}&&\beta E_{{\rm ex},0}=-\frac{\kappa^3}{8\pi}e^{-\kappa d};\\
\label{p14}
\hspace{-4mm}&&\beta P_0=2 n_i-\frac{\kappa^3}{24\pi}e^{-\kappa d}+\frac{4\pi}{3}d^3 n_i^2\left\{2+\left(\frac{\ell_{\rm B}}{d}\right)^2e^{-2\kappa d}\right\}.\nonumber\\
\hspace{-4mm}
\eea
The DHLLs for the excess energy and pressure follow from the vanishing ion size limit of Eqs.~(\ref{p13})-(\ref{p14}) as
\be
\label{p15}
\lim_{d\to0}\beta E_{{\rm ex},0}=-\frac{\kappa^3}{8\pi};\hspace{5mm}\lim_{d\to0}\beta P_0=2 n_i-\frac{\kappa^3}{24\pi}.
\ee

The comparison of Eqs.~(\ref{p13})-(\ref{p14}) with Eq.~(\ref{p15}) shows that at the gaussian level, the first type of ion size correction incorporated by the exponential factors is the attenuation of the opposite charge attraction by the minimum approach distance $d$. The resulting effect is illustrated in Fig.~\ref{fig1}(a) displaying the excess energy of a monovalent electrolyte. The comparison of the HNC data (red triangles) with the CCDH predictions~(\ref{p9}) (solid curve) and~(\ref{p11}) (blue disks) shows that the corresponding correction extends the accuracy of the DH-level energy (dashed curve) from $ n_i\sim0.01$ M up to $ n_i\sim0.3$ M. Beyond this concentration, the attenuation effect exaggerated by our CCDH formalism leads to the underestimation of the attractive excess energy.

The second HC correction to the DH formalism is the gaussian-level ionic volume fraction corresponding to the curly bracket term in the pressure identity~(\ref{p14}). In Eq.~(\ref{p12}), this contribution is augmented by the cumulant HC correction corresponding to the second term on the r.h.s.. At large packing fractions, these repulsive terms quadratic and cubic in the ion density bring the dominant contribution to pressure. This effect is illustrated in Figs.~\ref{fig1}(b)-(c) displaying the osmotic coefficient of monovalent electrolytes for two different ion sizes located in the regime of typical hydration radii. One sees that the numerical pressure result from Eq.~(\ref{p10}) and the analytical formula~(\ref{p12}) agree equally well with the HNC data both in the electrostatic correlation-driven ($ n_i\uparrow\phi\downarrow$) and the HC-dominated concentration regimes ($ n_i\uparrow\phi\uparrow$). 

Hence, the inclusion of HC size extends the accuracy of the DH-level pressure from $ n_i\sim10^{-2}$ M up to the molar concentration regime. One also notes that the difference between the gaussian-level pressure~(\ref{p14}) (dotted curves) and the CCDH result~(\ref{p10}) (solid curves) additionally including the perturbative correction~(\ref{eq22II}) is minor. Thus, at submolar concentrations, the gaussian-level HC size effects bring the dominant correction to the DH theory.

In order to identify the actual validity limit of our pressure identities, in Fig.~\ref{fig2}(a), we compare them with the MC simulations of Ref.~\cite{Svensson} extending into molar concentrations mainly governed by pronounced HC correlations. Therein, one sees that while the accuracy of the gaussian-level pressure~(\ref{p14}) is limited to $ n_i\sim2$ M, the CCDH results~(\ref{p10}) and~(\ref{p12}) agree with the MC data up to $ n_i\sim4$ M.  Beyond this concentration, the repulsive HC interactions are underestimated by our formalism. 

The corresponding limitation originating from the cumulant treatment of the bare HC interactions is equally observed in Fig.~\ref{fig2}(b) confronting our predictions with the separate MC data of Ref.~\cite{NetzMC}. Therein, one notes that in accordance with Fig.~\ref{fig2}(a), the large density regime of the plot is characterized by the underestimation of the MC data by Eq.~(\ref{p10}). One also sees that due to the substantially larger ion size,  the validity limit of our formalism drops down to submolar concentrations. Indeed, Fig.~\ref{fig2}(c) shows that the reduction of the ion size down to the range of typical hydration radii rises the accuracy limit of our formalism back to molar concentrations.

We identify now the corresponding regime of ionic packing fraction where our cumulant treatment of HC size breaks down. To this aim, we compare the neutral particle limit of the CCDH-level osmotic coefficient~(\ref{p12}),
\be
\label{l1}
\phi_{\rm HC}=\lim_{e\to0}\phi=1+4\eta+10t\eta^2,
\ee
where $\eta=\pi n_id^3/3$ is the packing fraction, and the virial expansion of the Carnahan-Starling (CS) pressure~\cite{CS},
\be
\label{l2}
\phi_{\rm HC}^{\rm (CS)}=\frac{1+\eta+\eta^2-\eta^3}{(1-\eta)^3}=1+4\eta+10\eta^2+18\eta^3+O\left(\eta^4\right).
\ee
Eqs.~(\ref{l1})-(\ref{l2}) show that our equation of state $\beta P=2n_i\phi$ correctly accounts for the HC interactions up to the cubic order in the salt density; the corresponding second virial coefficient $B_2=4$ in Eq.~(\ref{l1}) originates from the gaussian Hamiltonian~(\ref{eq22}), and the third virial coefficient $B_3=10$ is generated by the cumulant correction~(\ref{eq22II})~\cite{Hansen}.  The underestimation of the repulsive pressure in the HC-dominated regime of Figs.~\ref{fig2}(a)-(b) is precisely due to the higher order virial correction $18\eta^3$ absent in Eq.~(\ref{l1}). 

\begin{figure}
\includegraphics[width=1.\linewidth]{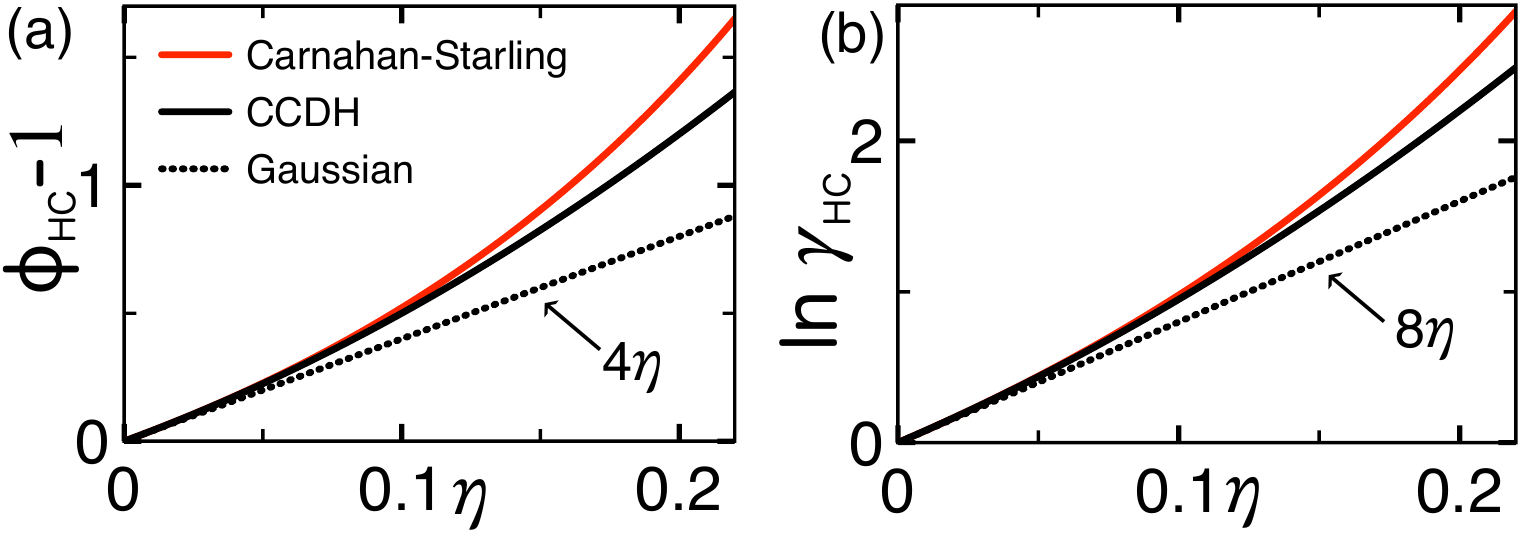}
\caption{(Color online) Neutral particle limit ($q_i\to0$) of (a) the osmotic coefficient and (b) the logarithm of the activity coefficient from the CCDH Eqs.~(\ref{l1}) and~(\ref{l3}) (solid black curves), the gaussian component of the latter (dotted curves), and the exact CS laws in Eqs.~(\ref{l2}) and~(\ref{l4}) (red curves).}  
\label{fig3}
\end{figure}

The effect of this limitation is  displayed in Fig.~\ref{fig3}(a). In agreement with Fig.~\ref{fig2}(b), one sees that the gaussian component of the HC pressure~(\ref{l1}) (dotted curve) and the full cumulant result (solid black curve) remain accurate roughly up to $\eta_{\rm c}\sim0.04$ and $\eta_{\rm c}\sim0.16$, respectively~\cite{rem3}.

Fig.~\ref{fig2}(d) shows that upon the further decrease of the HC size down to $d=2.38$ {\AA}, the CCDH pressure now overestimates the MC result. The overestimation of the MC pressure for small ions stems from the fact that at short interionic distances, the significant strength of the opposite charge attraction enhances the contribution of the attractive electrostatic energy~(\ref{p9}) to the pressure~(\ref{p10}). Moreover, in Fig.~\ref{fig1}(a), it was shown that our cumulant approach overestimating the size-induced attenuation of these attractions underestimates the electrostatic energy.  As a result, in Fig.~\ref{fig2}(d), the CCDH pressure exceeds the MC data. This said, at these particularly large salt concentrations, the ability of our formalism to reproduce qualitatively the general trend and the magnitude of the MC pressure is noteworthy.

Finally, in Fig.~\ref{fig1}(d), we consider divalent liquids mainly dominated by charge correlations. One sees that while the CCDH pressure is significantly more accurate than the DH prediction sharply diverging from the HNC data, the improved quantitative precision of our formalism is more limited than in the previously considered case of 1:1 electrolytes. Indeed, the inset shows that our cumulant approach mildly exaggerates the magnitude of the opposite charge attraction. This results in the underestimation of the osmotic pressure at low concentrations.

\subsection{Ionic activity coefficients}

\begin{figure*}
\includegraphics[width=1.\linewidth]{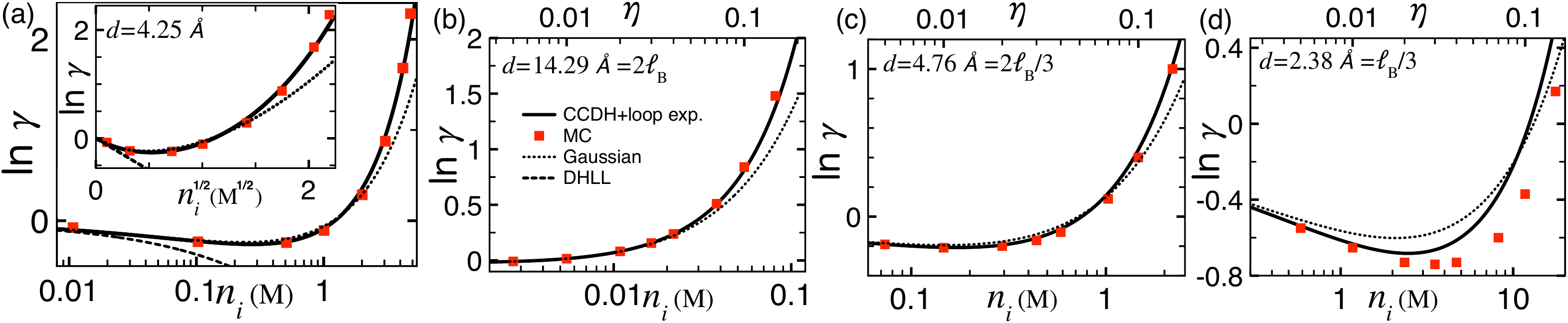}
\caption{(Color online) Logarithm of the activity coefficient of 1:1 solutions at various HC sizes. Solid curves: loop-expanded CCDH result~(\ref{co2}). Dotted curves: gaussian prediction~(\ref{co3}). Dashed curve in (a): DHLL. Symbols: MC data from (a) Fig. 2 of Ref.~\cite{Valleau} and (b)-(d) Table IV of Ref.~\cite{NetzMC} at the temperature $T=298$ K and dielectric constant $\e_{\rm w}=78.5$~\cite{rem1}.} 
\label{fig4}
\end{figure*}

The ionic activity of charged solutions is critically important for the characterization of the liquid structure at the molecular level. This macroscopic observable measuring the deviation of the liquid thermodynamics from the ideal gas behavior gives experimental access to the nature of the intermolecular interactions. In the case of monovalent solutions, the logarithm of the activity coefficient corresponds to the excess chemical potential. The latter can be related to the osmotic coefficient~(\ref{p10}) via the Gibbs-Duhem identity~\cite{book} as
\be
\label{co1}
\ln\gamma=\beta\mu_{\rm ex}=\phi-1+2\int_0^\kappa\frac{\mathrm{d}\kappa'}{\kappa'}\left(\phi-1\right).
\ee

For monovalent salt, the loop-expanded formula~(\ref{p12}) has been shown to be an accurate approximation of the osmotic coefficient~(\ref{p10}). Thus, we will use this analytical identity to calculate the ionic activity coefficient. Plugging Eq.~(\ref{p12}) into Eq.~(\ref{co1}), and evaluating the integral, the logarithm of the activity coefficient follows as
\begin{widetext} 
\bea
\label{co2}
\ln\gamma&=&\ln\gamma_0+15t\left(\frac{\pi}{3}n_id^3\right)^2\\
&&+\frac{t}{2304}\left\{-16\left[1+6\;{\rm E}_1(2\kappa d)+6\;{\rm Ei}(-\kappa d)+6\ln2\right]+8\left[8-\kappa d\left(4+(5\kappa d-19)\kappa d\right)\right]e^{-\kappa d}\right.\nonumber\\
&&\hspace{1.3cm}\left.+\left[3+12(1-2\kappa d)\kappa d\right]e^{-4\kappa d}+
\left[-51+2\kappa d\left(45+\kappa d\left(21+2\kappa d(5\kappa d-17)\right)\right)\right]e^{-2\kappa d}\right\}\nonumber\\
&&+\frac{t\Gamma}{6912\kappa d}\left\{-109+72\left[22+\kappa d\left(-17+2\kappa d(\kappa d-3)\right)\right]e^{-\kappa d}
-72\left[23+2\kappa d\left(-1+\kappa d\left(-4+(\kappa d-4)\kappa d\right)\right)\right]e^{-2\kappa d}\right.\nonumber\\
&&\left.\hspace{1.7cm}-8\left[-26+3\kappa d+36\left(\kappa d\right)^2\right]e^{-3\kappa d}-9\left[3+4\kappa d\left(3+2\kappa d(4\kappa d-7)\right)\right]e^{-4\kappa d}\right\}+O\left(\Gamma^2\right),\nonumber
\eea
\end{widetext}
where we defined the gaussian-level activity coefficient
\bea
\label{co3}
\ln\gamma_0&=&\frac{8\pi}{3} n_id^3\\
&&-\frac{\Gamma}{24\kappa d}\left\{7+4(\kappa d-2)e^{-\kappa d}\right.\nonumber\\
&&\left.\hspace{1.3cm}-\left[2\left(\kappa d\right)^2-2\kappa d-1\right]e^{-2\kappa d}\right\},\nonumber
\eea
and used the exponential integrals ${\rm E}_1(x)$ and ${\rm Ei}(x)$~\cite{math}.

Eq.~(\ref{co3}) indicates that the departure of the activity coefficient from ideality ($\gamma_0=1$) is set by the competition between the positive ionic volume fraction term embodying the HC repulsion, whose cumulant correction corresponds to the second term on the r.h.s. of Eq.~(\ref{co2}), and the negative term incorporating the attractive electrostatic correlations attenuated by the ion size. In the limit $d\to0$ where these ion size effects disappear, the DHLL follows from Eq.~(\ref{co2}) or~(\ref{co3}) as $\ln\gamma_{\rm DH}=-q_i^2\kappa\ell_{\rm B}/2$.

\begin{figure*}
\includegraphics[width=1.\linewidth]{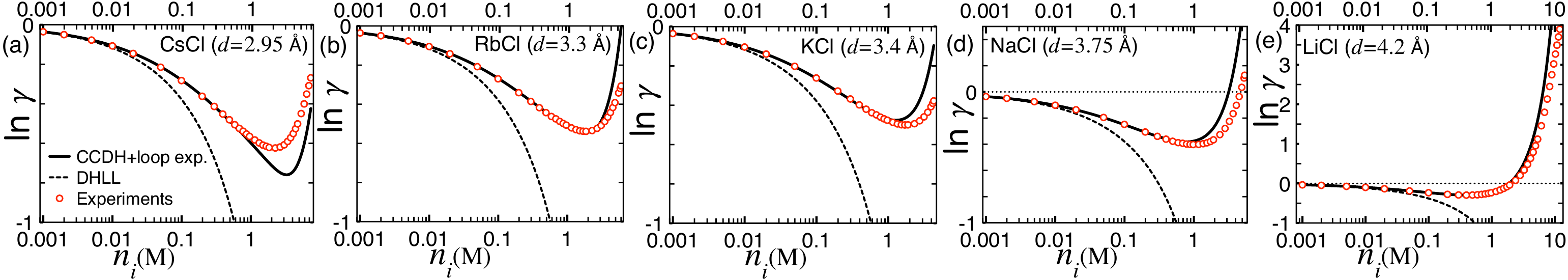}
\caption{(Color online) Logarithm of the ionic activity coefficient for various 1:1 solutions. The red circles are experimental data~\cite{rem2} from Ref.~\cite{Exp}, the dashed curves correspond to the DHLL $\ln\gamma_{\rm DH}=-q_i^2\kappa\ell_{\rm B}/2$, and the solid curves are the loop-expanded CCDH prediction~(\ref{co2}). The temperature is $T=298$ K. The adjusted HC sizes are indicated in the legends.} 
\label{fig5}
\end{figure*}

In Figs.~\ref{fig4}(a)-(c), one sees that over a broad range of ion sizes, the CCDH-level activity coefficient~(\ref{co2}) (solid curves) exhibits a good quantitative agreement with the MC data of Refs.~\cite{Valleau} and~\cite{NetzMC}. One also notes that the gaussian-level prediction~(\ref{co3}) (dotted curves) provides an accurate approximation of the full CCDH result~(\ref{co2}). More specifically, Fig.~\ref{fig4}(a) shows that in the regime of typical hydrated ion sizes, the accuracy of the DH law (dashed curve) limited to $ n_i\sim10$ mM is extended up to $ n_i\sim2$ M by the gaussian result, and to $ n_i\sim5$ M by the full CCDH prediction. Even in the small ion size regime of Fig.~\ref{fig4}(d) where our formalism has been shown to overestimate the size-induced attenuation of opposite charge attraction, the gaussian and CCDH results can reproduce qualitatively the electrostatically driven ($ n_i\uparrow\gamma\downarrow$) and the HC-dominated ($ n_i\uparrow\gamma\uparrow$) branches of the activity coefficient, and also predict the turnover concentration between these two regimes with reasonable accuracy.

In order to identify the accuracy level of our ionic activity identity~(\ref{co2}) in the consideration of the pure HC correlations, we compare its neutral HC sphere limit
\be
\label{l3}
\ln \gamma_{\rm HC}=\lim_{e\to0}\ln \gamma=8\eta+15t\eta^2
\ee
with the virial expansion of the CS chemical potential~\cite{Hansen} obtained by inserting Eq.~(\ref{l2}) into Eq.~(\ref{co1}) as
\be
\label{l4}
\ln \gamma_{\rm HC}^{\rm (CS)}=\frac{8\eta-9\eta^2+3\eta^3}{(1-\eta)^3}=8\eta+15\eta^2+24\eta^3+O\left(\eta^4\right).
\ee
In consistency with the pressure analysis of Sec.~\ref{epr}, Eqs.~(\ref{l3})-(\ref{l4}) show that HC interactions are correctly incorporated by the CCDH-level activity coefficient~(\ref{co2}) up to the quadratic order in the salt density. Fig.~\ref{fig3}(b) shows that this sets the validity limit of the gaussian and cumulant-level activities to $\eta_{\rm c}\sim0.05$ and $\eta_{\rm c}\sim0.18$, respectively~\cite{rem3}. Beyond the latter limit, the positive term $24\eta^3$ in Eq.~(\ref{l4}) missing in Eq.~(\ref{l3}) causes the underestimation of the ion activity by the CCDH formalism. This effect is equally displayed in Figs.~\ref{fig4}(a)-(b) by the MC data with the largest concentration. Finally, one notes that the agreement between the MC data and the theoretical ion activities in Fig.~\ref{fig4} is slightly better than that observed in the pressure plots of Fig.~\ref{fig2}. This stems from the fact that the virial expansion of the chemical potential~(\ref{l4}) converges faster than its pressure counterpart~(\ref{l2}) (compare the critical packing values $\eta_{\rm c}$ above).

\subsection{Confrontation with experiments}

Finally, we confront our prediction~(\ref{co2}) with experimental ionic activity data. At the large ion concentrations considered in our work, the dielectric response of the liquid is strongly suppressed by added salt~\cite{Sridhar1990,Wei1992,Buchner1999}. This dielectric decrement effect has been incorporated into our formalism via the replacement of the pure water permittivity $\e_{\rm w}$ in our equations with the salt-dependent permittivity $\e_{\rm el}( n_i)$ of the Gavish-Promislow model~\cite{Gavish}. In Appendix~\ref{dieldec}, we review the technical details of this model, and explain the fitting of its parameters by comparison with experimental dielectric permittivity data.

Fig.~\ref{fig5} compares the ionic activity formula~(\ref{co2}) and the DHLL with the experimental activity coefficients of five different monovalent electrolytes~\cite{rem2}. In these plots, the HC sizes have been adjusted to obtain the best agreement with the low density branch of the experimental data. The figure shows that the CCDH result can reproduce with reasonable accuracy the non-monotonic salt dependence of the experimental ionic activities. Namely, Eq.~(\ref{co2}) captures well beyond the DHLL the non-uniform slope of the experimental data in the ionic correlation-driven regime ($ n_i\uparrow\gamma\downarrow$) extending up to $n_i\sim1$ M, and reproduces qualitatively the reversal point and the subsequent rising trend in the HC-driven molar concentration regime ($ n_i\uparrow\gamma\uparrow$). It is also noteworthy that the adjusted ionic radii obey the cationic branch of the Hofmeister series, i.e.  ${\rm Cs}^+<{\rm Rb}^+<{\rm K}^+<{\rm Na}^+<{\rm Li}^+$~\cite{Hof1,Hof2}.

Interestingly, the molar concentration regime of Fig.~\ref{fig5} shows that the experimental data is underestimated by the CCDH prediction at small ion sizes and overestimated by our formalism at large ion sizes. We note that this quantitative disagreement is opposite to the departure of our predictions from the MC results in Fig.~\ref{fig4}. Thus, at molar concentrations, the deviation of our ionic activity curves from the experimental  data may originate mainly from model limitations rather than our approximate treatment of the many-body ion interactions.  Indeed, at these large salt concentrations where the volume fraction of the hydrated ions becomes comparable with that of the free solvent molecules, explicit solvent effects of HC and electrostatic origin are expected to play a sizeable role. The inclusion of explicit solvent into our model is needed to shed light on this issue~\cite{BuyukDielDec}.

\section{Conclusions}

In this work, via the cumulant expansion of the electrostatic SD equations, we carried out the systematic incorporation of the HC ion size into the DH formalism of bulk electrolytes. By confronting the predictions of the CCDH theory with a large variety of MC and HNC data from the literature~\cite{NetzMC,Svensson,Valleau,AttardPRE}, we showed that this upgrade boost the accuracy of the DHLLs for the osmotic pressure and the activity coefficient of monovalent ions from submolar into molar salt concentrations. This significant extension is the main result of the present work.

By carrying out the loop expansion of the CCDH formulas, we derived as well analytically insightful pressure and ionic activity identities that may be useful to experimentalists. Finally, the activity coefficients of the CCDH formalism augmented by the effect of dielectric decrement have been confronted with experimental data. It was shown that for five different types of monovalent electrolytes, via the only adjustment of the hydrated ion size, the present formalism can reproduce with reasonable accuracy the non-uniform salt dependence of the experimental ionic activity coefficients up to several molar concentrations. The inequality satisfied by these HC diameters was found to obey the Hofmeister series~\cite{Hof1,Hof2}. 

The limitations of the calculation scheme underlying our formalism have been meticulously assessed. The first limitation of our approach originates from the cumulant treatment of the bare HC interactions equivalent to the second order virial expansion of the CS identities.  In the regime of large concentrations, this cumulant approach (i) underestimates the interionic HC repulsion, and (ii) overestimates the HC size-induced attenuation of opposite charge attraction. In Fig.~\ref{fig2}(a), the cancellation of these errors may be indeed responsible for the extension of the agreement between the CCDH- and MC-level pressures beyond the critical packing fraction $\eta\ce\sim0.16$ or concentration $n_{i,{\rm c}}\sim3.3$ M where the neutral particle limit of the CCDH pressure displayed in Fig.~\ref{fig3}(a) has been shown to loose its quantitative validity.

In the case of divalent ions mainly governed by charge correlations, the improved precision of our formalism was shown to be limited to submolar salt concentrations. This additional limitation is caused by the purely quadratic dependence of the reference Hamiltonian~(\ref{eq22}) on the electrostatic potential $\psi\ce(\br)$. Indeed, in the presence of strongly coupled multivalent charges, this WC assumption neglecting the non-linearities in charge of suppressing the potential fluctuations leads to the overestimation of the opposite charge attraction.

The field theory framework underlying the CCDH formalism offers potential for future improvements. First of all, the field theoretic formulation of electrolyte solutions is known to be well-suited for the application of systematic perturbation techniques and self-consistent calculation schemes~\cite{NetzLoop,NetzSC,NetzVir,netzvar,HatloVar,Buyuk2020,Podgornik2010}. Indeed, in future works, the aforementioned limitations can be relaxed by the extension of the underlying cumulant expansion  to higher orders, and via the incorporation of electrostatic interactions by more sophisticated computational techniques bypassing the WC gaussian approximation.

Additionally, as discussed in the Introduction, the field theory approach to charged systems has proved to be an efficient way to incorporate various relevant details of electrolytes, such as the intramolecular ion structure~\cite{Demery,Buyuk2023}, explicit solvent~\cite{DuncSolv,NL1}, inhomogeneous liquid partition in nanofludic charge transport~\cite{JPCB2020} and polymer translocation through nanopores~\cite{BuyukLang2022}, and conformational polymer fluctuations in salty polymeric solutions~\cite{NetzPol,RudiPol}. The application of the CCDH theory to the aforementioned systems can extend the quantitative validity of the predictions by purely electrostatic formalisms into the regime of molar salt concentrations. Finally, within the present formalism, we currently investigate the effect of ion correlations on the screening length and the dielectric permittivity of bulk electrolytes~\cite{AttardPRE,AttardJCP,Kj1,Kj2,KjJCP2016,Kj2020,NetzMC}. This ongoing work will be reported in a follow-up article.

\smallskip
\textbf{Supporting Information}

Derivation of the excess energy density and equation of state, salt dependent dielectric permittivities.

\smallskip
\textbf{ACKNOWLEDGMENTS}

We would like to thank R. R. Netz for providing many useful suggestions, and for stressing the importance of confronting our theoretical predictions with experimental ionic activity data by incorporating the effect of salt-induced dielectric decrement into our formalism.

The author received no financial support for this work.

\smallskip
\appendix

\section{Derivation of the excess energy density and equation of state}

\label{ap0}

In this appendix, we review the derivation of the excess energy density and the osmotic pressure~\cite{Hansen} for mixed electrolytes of general composition.

\subsection{Excess energy density}
\label{ap1}

The excess energy density corresponds to the statistical average of the pairwise interaction energy~(\ref{eq2}) per volume. Making use of the density representation in Eq.~(\ref{eq8}), this average can be expressed as
\bea\label{a1}
&&\beta E_{\rm ex}=\frac{1}{V}\lan\beta E\m\ran\G\\
&&=\frac{1}{2V}\int\mathrm{d^3}\br\mathrm{d^3}\br'\left[\lan\hn\ce(\br)v\ce(\br-\br')\hn\ce(\br')\ran\G\right.\nonumber\\
&&\left.\hspace{2.6cm}+\lan\hn\h(\br)v\h(\br-\br')\hn\h(\br')\ran\G\right]\nonumber\\
&&=\int\frac{\mathrm{d^3}\br\mathrm{d^3}\br'}{2V}\sum_{i=1}^p\sum_{j=1}^p\left[v\h(\br-\br')+q_iq_jv\ce(\br-\br')\right]\nonumber\\
&&\hspace{2.7cm}\times\lan\sum_{k=1}^{N_i}\sum_{l=1}^{N_j}\delta(\br-\br_{ik})\delta(\br'-\br_{jl})\ran\G,\nonumber
\eea
where $V$ is the total volume, and $\lan\cdot\ran_{\rm G}$ denotes the grand-canonical average. Using the definition of the pair distribution function~(\ref{p1}) together with the translational invariance in the bulk liquid, Eq.~(\ref{a1}) can be recast as
\be\label{a2}
\beta E_{\rm ex}=\frac{1}{2}\sum_{i=1}^p\sum_{j=1}^p n_i n_j\int_0^\infty\mathrm{d^3\br}\left[v\h(\br)+q_iq_jv\ce(\br)\right]g_{ij}(\br).
\ee

Using now the identity $g_{ij}(\br)=g_{ij}(r)=H_{ij}(r)+1$, as well as Eqs.~(\ref{eq5}) and~(\ref{p5}), one finds that the HC component of the integral in Eq.~(\ref{a2}) vanishes, i.e.
\bea
\label{a3}
&&\int_0^\infty\mathrm{d}rr^2v\h(r)g_{ij}(r)\\
&&=\int_d^\infty\mathrm{d}rr^2v\h(r)\left[1+T_{ij}(r)\right]e^{-q_iq_jG_0(r)}=0,\nonumber
\eea
where we accounted for the cancellation of the HC potential outside the contact sphere, i.e. $v\h(r>d)=0$. Finally, taking into account the electroneutrality condition~(\ref{eq33}), the excess energy density~(\ref{a2}) becomes
\be\label{a4}
\beta E_{\rm ex}=2\pi\ell_{\rm B}\sum_{i=1}^p\sum_{j=1}^p n_i n_jq_iq_j\int_d^\infty\mathrm{d}rrH_{ij}(r).
\ee
For symmetric electrolytes, Eq.~(\ref{a4}) reduces to
\be
\label{a4II}
\beta E_{\rm ex}=-4\pi\ell_{\rm B}q_i^2 n_i^2\int_d^\infty\mathrm{d}rr\left[H_{+-}(r)-H_{++}(r)\right].
\ee

\subsection{Osmotic pressure}
\label{ap2}

The derivation of the equation of state is based on the virial theorem~\cite{Hansen}
\bea
\label{a5}
\beta P=\sum^p_{i=1} n_i-\frac{S}{3V},
\eea
with the averaged term defined as
\be\label{a6}
S=\lan\sum_{i=1}^p\sum_{j=1}^{N_i}\br_{ij}\cdot\nabla_{ij}\left(\beta E\m\right)\ran\G,
\ee
where $\nabla_{ij}$ is the derivative with respect to the ionic position vector $\br_{ij}$. Switching now from energy to force, and using the superposition principle, Eq.~(\ref{a6}) becomes
\bea
\label{a7}
\hspace{-7mm}S&=&-\lan\sum_{i=1}^p\sum_{j=1}^{N_i}\br_{ij}\cdot\sum_{k=1}^p\sum_{l=1}^{N_k}\F_{kl,ij}\ran\G\\
\label{a8}
\hspace{-7mm}&=&-\frac{1}{2}\lan\sum_{i=1}^p\sum_{j=1}^{N_i}\sum_{k=1}^p\sum_{l=1}^{N_k}\left(\br_{ij}-\br_{kl}\right)\cdot\F_{kl,ij}\ran\G,
\eea
where $\F_{kl,ij}$ is the force exerted by the ion $l$ of the species $k$ on the ion $j$ of the species $i$. In order to pass from Eq.~(\ref{a7}) to Eq.~(\ref{a8}), we used Newton's third law, and permuted the summation indices. At this point, we switch back to the potential representation. This yields
\begin{widetext}
\bea
\label{a9}
S&=&\frac{1}{2}\lan\sum_{i=1}^p\sum_{j=1}^{N_i}\sum_{k=1}^p\sum_{l=1}^{N_k}||\br_{ij}-\br_{kl}||\left\{v'\h\left(||\br_{ij}-\br_{kl}||\right)+q_iq_kv'\ce\left(||\br_{ij}-\br_{kl}||\right)\right\}\ran\G\\
\label{a10}
&=&\frac{1}{2}\int\mathrm{d}^3\br\mathrm{d}^3\br'||\br-\br'||\sum_{i=1}^p\sum_{k=1}^p\left\{v'\h\left(||\br-\br'||\right)+q_iq_kv'\ce\left(||\br-\br'||\right)\right\}
\lan\sum_{j=1}^{N_i}\sum_{l=1}^{N_k}\delta(\br-\br_{ij})\delta(\br'-\br_{kl})\ran\G.
\eea
\end{widetext}
In Eqs.~(\ref{a9})-(\ref{a10}), the prime sign denotes the derivative of the function with respect to its argument. Using now Eq.~(\ref{p1}), and accounting for the translational invariance in the system, Eq.~(\ref{a10}) can be expressed in terms of the pair distribution function as
\be
\label{a11}
S=2\pi V\int_0^\infty\mathrm{d}rr^3\sum_{i=1}^p\sum_{k=1}^p n_i n_k\left\{v'\h(r)+q_iq_kv'\ce(r)\right\}g_{ik}(r).
\ee

At this point, we use the identity $g_{ij}(r)=H_{ij}(r)+1$ together with Eqs.~(\ref{eq5}) and~(\ref{p5}) to simplify the HC component of Eq.~(\ref{a11}) as follows:
\bea
\label{a12}
\int_0^\infty\mathrm{d}rr^3v'\h(r)g_{ik}(r)&=&-\int_0^\infty\mathrm{d}rr^3\left[1+T_{ik}(r)\right]\\
&&\hspace{1.1cm}\times e^{-q_iq_kG_0(r)}\partial_r H(r-d)\nonumber\\
\label{a13}
&=&-d^3\left[H_{ik}(d^+)+1\right].
\eea
In order to pass from Eq.~(\ref{a12}) to~(\ref{a13}), we used the identity $H'(x)=\delta(x)$. Plugging Eq.~(\ref{a13}) into Eq.~(\ref{a11}), and taking into account the electroneutrality condition~(\ref{eq33}), the osmotic pressure~(\ref{a5}) finally becomes
\bea
\label{a14}
\hspace{-7mm}\beta P&=&\sum^p_{i=1} n_i+\frac{2\pi}{3}d^3\sum_{i=1}^p\sum_{k=1}^p n_i n_k\left[H_{ik}(d^+)+1\right]\\
\hspace{-7mm}&&+\frac{2\pi\ell_{\rm B}}{3}\sum_{i=1}^p\sum_{k=1}^p n_i n_kq_iq_k\int_d^\infty\mathrm{d}rrH_{ik}(r).\nonumber
\eea
In the case of symmetric solutions, Eq~(\ref{a14}) reduces to
\be
\label{a14II}
\beta P=2 n_i+\frac{4\pi}{3}d^3 n_i^2\left[H_{+-}(d^+)+H_{++}(d^+)+2\right]+\frac{\beta E_{\rm ex}}{3}.
\ee

\subsection{Loop expansion of the thermodynamic identities~(\ref{a4II}) and ~(\ref{a14II})}
\label{ap3}

In order to derive closed-form expressions for the excess energy~(\ref{a4II}) and the equation of state~(\ref{a14II}), we carry out here the loop expansion of these identities for symmetric $q_i:q_i$ electrolytes. This requires first the loop expansion of the pair correlation functions defined by Eq.~(\ref{p5}). To this aim, we note that the components of the Fourier-expanded auxiliary function~(\ref{p7}) are given by
\bea
\label{a15}
T_{++}(r)&=&\int_0^\infty\frac{\mathrm{d}qq^2}{2\pi^2}\left\{ n_i\left[\tilde{h}^2_{++}(q)+\tilde{h}^2_{+-}(q)-2q_i^4\tilde G_0^2(q)\right]\right.\nonumber\\
&&\left.\hspace{1.7cm}-q_i^2\tilde G_1(q)\right\}\frac{\sin(qr)}{qr};\\
\label{a16}
T_{+-}(r)&=&\int_0^\infty\frac{\mathrm{d}qq^2}{2\pi^2}\left\{2 n_i\left[\tilde{h}_{++}(q)\tilde{h}_{+-}(q)+q_i^4\tilde G_0^2(q)\right]\right.\nonumber\\
&&\left.\hspace{1.7cm}+q_i^2\tilde G_1(q)\right\}\frac{\sin(qr)}{qr},
\eea
where the FT of the cumulant correction~(\ref{eq39II}) to the Green's function reads
\be
\label{a17}
\tilde G_1(q)=-2 n_i^2q_i^2\tilde G_0^2(q)\left[\tilde{h}_{++}(q)-\tilde{h}_{+-}(q)+2q_i^2\tilde G_0(q)\right].
\ee
At this point, we introduce the dimensionless lengths $\bar{r}=\kappa r$ and $\bar{d}=\kappa d$, and the dimensionless wavevector $\bar{q}=q/\kappa$. In terms of these variables, the functions~(\ref{a15})-(\ref{a16}) become
\begin{widetext}
\bea
\label{a18}
T_{++}(\bar{r})&=&\int_0^\infty\frac{\mathrm{d}\bar{q}\bar{q}^2}{16\pi^3\Gamma}\left\{\bar{h}^2_{++}(\bar{q})+\bar{h}^2_{+-}(\bar{q})-2\Gamma^2\bar{u}^2(\bar{q})
+\frac{\Gamma}{4\pi}\bar{u}^2(\bar{q})\left[\bar{h}_{++}(\bar{q})-\bar{h}_{+-}(\bar{q})+2\Gamma \bar{u}(\bar{q})\right]\right\}\frac{\sin(\bar{q}\bar{r})}{\bar{q}\bar{r}};\\
\label{a19}
T_{+-}(\bar{r})&=&\int_0^\infty\frac{\mathrm{d}\bar{q}\bar{q}^2}{16\pi^3\Gamma}\left\{2\bar{h}_{++}(\bar{q})\bar{h}_{+-}(\bar{q})+2\Gamma^2\bar{u}^2(\bar{q})
-\frac{\Gamma}{4\pi}\bar{u}^2(\bar{q})\left[\bar{h}_{++}(\bar{q})-\bar{h}_{+-}(\bar{q})+2\Gamma \bar{u}(\bar{q})\right]\right\}\frac{\sin(\bar{q}\bar{r})}{\bar{q}\bar{r}},
\eea
where we defined the rescaled FT of the DH potential, $\bar{u}(\bar{q})=4\pi/(\bar{q}^2+1)$, the electrostatic coupling parameter $\Gamma=q_i^2\ell_{\rm B}\kappa$, and the rescaled FT of the Mayer function~(\ref{eq43}),
\be\label{a20}
\bar{h}_{+\pm}(\bar{q})=-\frac{4\pi}{\bar{q}^3}\left[\sin(\bar{q}\bar{d})-\bar{q}\bar{d}\cos(\bar{q}\bar{d})\right]+\frac{4\pi}{\bar{q}}\int_{\bar{d}}^\infty\mathrm{d}\bar{r}\bar{r}\sin(\bar{q}\bar{r})\left\{e^{\mp\Gamma u(\bar{r})}-1\right\},
\ee
with the dimensionless DH potential $u(\bar{r})=e^{-\bar{r}}/\bar{r}$. 

Now, we Taylor-expand Eq.~(\ref{a20}) as
\be\label{a21}
\bar{h}_{+\pm}(\bar{q})=f_0(\bar{q})\mp\Gamma f_1((\bar{q}))+\frac{\Gamma^2}{2}f_2((\bar{q}))+O\left(\Gamma^3\right),
\ee
with the auxiliary functions defined as
\be\label{a22}
f_0(\bar{q})=-\frac{4\pi}{\bar{q}^3}\left[\sin(\bar{q}\bar{d})-\bar{q}\bar{d}\cos(\bar{q}\bar{d})\right];\hspace{5mm}f_n(\bar{q})=\frac{4\pi}{\bar{q}}\int_{\bar{d}}^\infty\mathrm{d}\bar{r}\bar{r}\sin(\bar{q}\bar{r})u^n(\bar{r}).
\ee
Plugging the expansion~(\ref{a21}) into Eqs.~(\ref{a18})-(\ref{a19}), the loop expansions of the latter become
\be
\label{a23}
T_{+\pm}(\bar{r})=\int_0^\infty\frac{\mathrm{d}\bar{q}\bar{q}^2}{8\pi^3\Gamma}\left\{f^2_0(\bar{q})\pm\Gamma^2\left[f_1^2(\bar{q})+f_0(\bar{q})f_2(\bar{q})-\bar{u}^2(\bar{q})\right]
\mp\frac{\Gamma^2}{4\pi}\bar{u}^2(\bar{q})\left[f_1(\bar{q})-\bar{u}(\bar{q})\right]\right\}\frac{\sin(\bar{q}\bar{r})}{\bar{q}\bar{r}}+O\left(\Gamma^3\right).
\ee
Using the definitions in Eq.~(\ref{a22}), and switching the order of the spatial and Fourier integrals, the latter can be analytically evaluated, and Eq.~(\ref{a23}) takes the closed-form 
\be\label{a24}
T_{+\pm}(\bar{r})=\frac{1}{\Gamma}V(\bar{r})+\Gamma\left[\pm J_0(\bar{r})+J_1(\bar{r})\right]+O\left(\Gamma^3\right),
\ee
with the auxiliary functions
\bea\label{a25}
V(\bar{r})&=&\frac{1}{48}\left(\bar{r}-2\bar{d}\right)^2\left(\bar{r}+4\bar{d}\right)\theta\left(2\bar{d}-\bar{r}\right);\\
\label{a26}
J_0(\bar{r})&=&\frac{1}{2\bar{r}}\left\{(1-e^{-\bar{r}})e^{-2\bar{d}}\theta(2\bar{d}-\bar{r})+\left(\bar{r}-2\bar{d}+1-e^{-2\bar{d}}\right)e^{-\bar{r}}\theta(\bar{r}-2\bar{d})\right\}\\
&&+\frac{e^{-\bar{r}}}{16\bar{r}}\left\{(3+2\bar{r}+2\bar{d})e^{-2\bar{d}}+4\bar{d}(1+\bar{r})-10\bar{r}-2\bar{d}^2-3\right\};\nonumber\\
\label{a27}
J_1(\bar{r})&=&\frac{1}{16\bar{r}}\left\{\left[2\bar{r}-2\bar{d}-1-(2\bar{r}+2\bar{d}-1)e^{4\bar{d}}\right]e^{-2(\bar{r}+\bar{d})}+4(\bar{r}^2-\bar{d}^2)\left[\mathrm{Ei}(-2\bar{d}-2\bar{r})-\mathrm{Ei}(2\bar{d}-2\bar{r})\right]\right\}\theta\left(\bar{r}-2\bar{d}\right)\nonumber\\
&&+\frac{1}{16\bar{r}}\left\{\left(2\bar{r}-2\bar{d}-1\right)e^{-2(\bar{r}+\bar{d})}+(1-4\bar{r}+2\bar{d})e^{-2\bar{d}}+4(\bar{d}^2-\bar{r}^2)\left[\mathrm{Ei}(-2\bar{d})-\mathrm{Ei}(-2\bar{d}-2\bar{r})\right]\right\}\theta\left(2\bar{d}-\bar{r}\right).\nonumber
\eea
In Eqs.~(\ref{a25})-(\ref{a27}), we used the Heaviside step function $\theta(x)$ and the exponential integral function ${\rm Ei}(x)$~\cite{math}.

In terms of the dimensionless parameters defined above, the pair correlation function~(\ref{p5}) reads
\be
\label{a28II}
H_{+\pm}(\bar{r})=\theta(\bar{r}-\bar{d})\left[1+tT_{+\pm}(\bar{r})\right]e^{\mp\frac{\Gamma}{\bar{r}} e^{-\bar{r}}}
\ee
Inserting Eq.~(\ref{a24}) into Eq.~(\ref{a28II}), and expanding the result in terms of the coupling parameter $\Gamma$, one obtains
\bea
\label{a28}
H_{+\pm}(\bar{r})&=&\theta(\bar{r}-\bar{d})\left\{1\mp\Gamma\frac{e^{-\bar{r}}}{\bar{r}}+\Gamma^2\frac{e^{-2\bar{r}}}{2\bar{r}^2}+\frac{t}{\Gamma}V(\bar{r})\mp tV(\bar{r})\frac{e^{-\bar{r}}}{\bar{r}}+t\Gamma\left[\pm J_0(\bar{r})+J_1(\bar{r})+V(\bar{r})\frac{e^{-2\bar{r}}}{2\bar{r}^2}\right]\right.\\
&&\left.\hspace{1.5cm}-t\Gamma^2\left[J_0(\bar{r})\pm J_1(\bar{r})\right]\frac{e^{-\bar{r}}}{\bar{r}}\right\}\nonumber\\
&&-1+O\left(\Gamma^3\right).
\nonumber
\eea

Substituting now the loop-expanded pair correlation functions in Eq.~(\ref{a28}) into Eqs.~(\ref{a4II}) and~(\ref{a14II}), and carrying out the spatial integrals, the excess energy per concentration and the equation of state follow in the closed-forms
\bea
\label{a29}
&&\frac{\beta E_{\rm ex}}{ n_i}=-\Gamma e^{-\bar{d}}+\frac{t}{48}\left\{6(1+2\bar{d})e^{-2\bar{d}}-(6+6\bar{d}-9\bar{d}^2+5\bar{d}^3)e^{-\bar{d}}\right\}\\
&&\hspace{1.3cm}+\frac{t\Gamma}{16}\left\{(4\bar{d}-3)e^{-3\bar{d}}+8(2+\bar{d})e^{-2\bar{d}}+(2\bar{d}^2-2\bar{d}-13)e^{-\bar{d}}\right\}+O\left(\Gamma^2\right);\nonumber\\
\label{a30}
&&\frac{\beta P}{2 n_i}=1+\frac{\bar{d}^3}{6\Gamma}+\frac{\Gamma}{6}\left[\frac{\bar{d}}{2}e^{-2\bar{d}}-e^{-\bar{d}}\right]+\frac{5t}{288\Gamma^2}\bar{d}^6\nonumber\\
&&\hspace{6mm}+\frac{t}{576}\left\{\left[5\bar{d}^4-12\bar{d}(e^{\bar{d}}-2)-12(e^{\bar{d}}-1)-2\bar{d}^3(5e^{\bar{d}}+6)\right]e^{-2\bar{d}}+6\bar{d}^2\left(3e^{-\bar{d}}+e^{-2\bar{d}}-e^{-4\bar{d}}\right)\right\}\nonumber\\
&&\hspace{6mm}+\frac{t\Gamma}{96}\left\{(5-4\bar{d})\bar{d}e^{-4\bar{d}}-(4\bar{d}+3)e^{-3\bar{d}}+\left(16+11\bar{d}+6\bar{d}^2-2\bar{d}^3\right)e^{-2\bar{d}}+\left(2\bar{d}^2-2\bar{d}-13\right)e^{-\bar{d}}\right\}+O\left(\Gamma^2\right).\nonumber
\eea
Finally, expressing Eqs.~(\ref{a29})-(\ref{a30}) in physical units, one obtains the identities~(\ref{p11})-(\ref{p12}) in the main text.

\end{widetext}

\begin{figure}
\includegraphics[width=1.\linewidth]{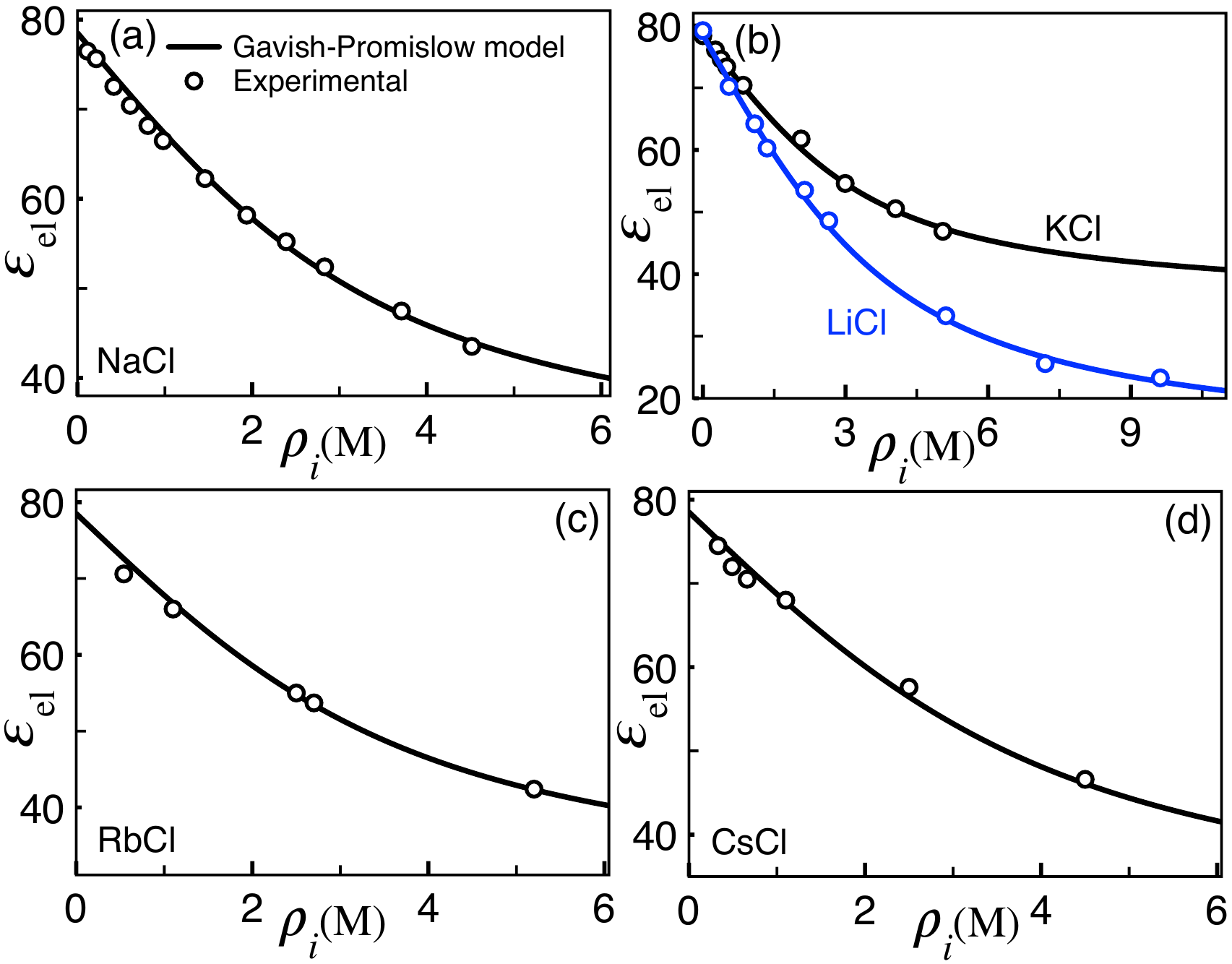}
\caption{(Color online) Experimental dielectric constants (symbols) and the Gavish-Promislow prediction~(\ref{dec1}) (solid curves) against salt concentration for various electrolytes.  The experimental data are from Ref.~\cite{Buchner1999} (NaCl),  Ref.~\cite{NetzMC2} (KCl), Ref.~\cite{Sridhar1990} (LiCl), and Ref.~\cite{Wei1992} (CsCl and RbCl).}  
\label{figApp}
\end{figure}

\section{Salt dependent dielectric permittivities}
\label{dieldec}

We report here the salt-dependent dielectric permittivity formula of the Gavish-Promislow model~\cite{Gavish}, which has been incorporated into our formalism for the comparison of the theoretical ionic activity curves with the experimental data in Fig.~\ref{fig5} of the main text. Within the framework of this model, the salt-dependent dielectric permittivity of the electrolyte is given by
\be
\label{dec1}
\e_{\rm el}=\e_{\rm w}+\left(\e_{\rm w}-\e_{\rm ms}\right){\mathcal L}\left(\frac{3\alpha n_i}{\e_{\rm w}-\e_{\rm ms}}\right),
\ee
where $\mathcal{L}(x)=\coth x-x^{-1}$ is the Langevin function, $\alpha$ stands for the ionic excess polarizability, and $\e_{\rm ms}$ is the dielectric permittivity of the molten salt. Below, we provide the numerical values of these parameters for the five type of electrolytes considered in the present work.

 Fig.~\ref{figApp} compares the salt-dependence of the permittivity function~(\ref{dec1}) and the experimental permittivity data whose references are provided in the caption. The model parameters taken from Ref.~\cite{NetzMC2} are $\e_{\rm ms}=27.9$ and $\alpha=-11.59$ ${\rm M}^{-1}$ for the NaCl solution, and $\e_{\rm ms}=35$ and $\alpha=-10.02$ ${\rm M}^{-1}$ for the KCl electrolyte. For the remaining electrolyte solutions, the parameters providing the best fit to the experimental data were determined by us as $\e_{\rm ms}=11$ and $\alpha=-13.5$ ${\rm M}^{-1}$ for LiCl,  $\e_{\rm ms}=27$ and $\alpha=-11$ ${\rm M}^{-1}$ for RbCl, and $\e_{\rm ms}=27$ and $\alpha=-10$ ${\rm M}^{-1}$ for CsCl.


\newpage
{\bf REFERENCES}
\begin{thebibliography}{99}

\bibitem {Dyson} Dyson; F. Ground‐State Energy of a Finite System of Charged Particles. {\it J. Math. Phys.} {\bf 1967}, {\it 8}, 1538–1545.
\bibitem {OppenAstro}  Oppenheimer, J. R.; Volkoff, G. M. On Massive Neutron Cores. {\it Phys. Rev.} {\bf 1939}, {\it 55}, 374-381.
\bibitem {Bont} Bonthuis, D. J.; Zhang, J.; Hornblower, B.; Math\'{e}, J.; Shklovskii, B. I.; Meller, A. Self-Energy-Limited Ion Transport in Subnanometer Channels. {\it Phys. Rev. Lett.} {\bf 2006}, {\it 97}, 128104. 
\bibitem {Holm} C. Holm, P. Kekicheff, and R. Podgornik, Electrostatic Effects in Soft Matter and Biophysics, Kluwer Academic, Dordrecht (2001).
\bibitem {gn1} Podgornik, R.; Strey, H. H.; Parsegian, V. A. \textit{Molecular Interactions in Lipids, DNA and DNA-lipid Complexes, in Gene Therapy: Therapeutic Mechanisms and Strategies}, 209-239, Marcel Dekker, New York, 2000.
\bibitem {Boc1} Sendner, C.; Horinek, D.; Bocquet, L.; Netz, R. Interfacial Water at Hydrophobic and Hydrophilic Surfaces: Slip, Viscosity, and Diffusion. \textit{Langmuir} {\bf 2009}, {\it 25}, 10768-10781.
\bibitem {Boc2} Bocquet, L. Nanofluidics Coming of Age. {\it Nature Materials} {\bf 2020}, {\it 19}, 254. 
\bibitem {rev2} Zhang, L. L.;  Zhao, X. S. Carbon-based materials as supercapacitor electrodes. {\it Chem. Soc. Rev.} {\bf 2009} {\it 38}, 2520-2531.
\bibitem {sf} Ram, V.; Salkuty, S. R. An Overview of Major Synthetic Fuels. {\it Energies} {\bf 2023}, {\it 16}, 2834.
\bibitem {pur} Teh, C. Y. ; Budiman, P. M.; Shak, K. P. Y.; Wu, T. Y. Recent Advancement of Coagulation–Flocculation and Its Application in Wastewater Treatment. {\it Ind. Eng. Chem. Res.} {\bf 2016}, {\it 55}, 4363-4389.
\bibitem {Yar} Yaroshchuk, A. Non-steric Mechanism of Nanofiltration: Superposition of Donnan and Dielectric Exclusion. {\it Sep. Purif. Technology} {\bf 2001}, {\it 143}, 22.
\bibitem {Szymczyk} Szymczyk, A.; Fievet, P. Investigating transport properties of nanofiltration membranes by means of a steric, electric and dielectric exclusion model. {\it J. Membrane Sci.} {\bf 2005} {\it 252}, 77-88.
\bibitem {Book} J.F., Zemaitis Jr., D.M., Clark; M, Rafal; N.C., Scrivner {\it Handbook of aqueous electrolyte thermodynamics: theory and application}; Wiley: 1986, New York.
\bibitem {DH} Debye, P.; H\"{u}ckel, E. Zur Theorie der Elektrolyte. {\it Phys. Z.} {\bf 1923}, {\it 24}, 185-206.
\bibitem {Valleau} Valleau, J.P.; Cohen, L.K. Primitive model electrolytes. I. Grand canonical Monte Carlo computations. {\it J. Chem. Phys.} {\bf 1980}, {\it 72}, 5935-5941. 
\bibitem {Svensson} Svensson, B.R.; Woodward, C.E. Widom's method for uniform and non-uniform electrolyte solutions. {\it Mol. Phys.} {\bf 1988}, {\it 64}, 247-259. 
\bibitem {NetzMC} Janecek, J.; Netz, R.R. Effective screening length and quasiuniversality for the restricted primitive model of an electrolyte solution. {\it J. Chem. Phys.} {\bf 2009}, {\it 130}, 074502. 
\bibitem {NetzMC2} P. dos Santos, A.; Uematsu. Y.; Rathert, A.; Loche, P.; and Netz, R. R. Consistent description of ion-specificity in bulk and at interfaces by solvent implicit simulations and mean-field theory. {\it J. Chem. Phys.} {\bf 2020}, {\it 153}, 034103.
\bibitem {LevinMC} Bakhshandeh, A.; Levin, Y.  Widom insertion method in simulations with Ewald summation. {\it J. Chem. Phys.} {\bf 2022}, {\it 156}, 134110.
\bibitem {Hansen} Hansen, J.P.; McDonald, I.R. {\it Theory of Simple Liquids}; 2nd edition, Academic Press: 1990. 
\bibitem {Blum} Blum, L. Mean spherical model for asymmetric electrolytes. {\it Mol. Phys.} {\bf 1975}, {\it 30}, 1529-1535. 
\bibitem {Henderson}  Henderson, D.; Smith, W. R. Exact analytical formulas for the distribution functions of charged hard spheres in the mean spherical approximation. {\it J. Stat. Phys.} {\bf 1978}, {\it 19}, 191-200.  
\bibitem {Boda} Gillespie, D.; Valisko, M.; Boda, D. Electrostatic correlations in electrolytes: Contribution of screening ion interactions to the excess chemical potential. {\it J. Chem. Phys.} {\bf 2021}, {\it 155}, 221102. 
\bibitem {Hoye} Hoye, J. S.; Lebowitz, J. L.; Stell, G. Generalized mean spherical approximations for polar and ionic fluids. {\it J. Chem. Phys.} {\bf 1974}, {\it 61}, 3253-3260. 
\bibitem {Roji} Cats, P.; Evans, R.; Hartel, A.; van Roji, R. Primitive model electrolytes in the near and far field: Decay lengths from DFT and simulations. {\it J. Chem. Phys.} {\bf 2021}, {\it 154}, 124504. 
\bibitem {AttardPRE} Attard, P.  Asymptotic analysis of primitive model electrolytes and the electrical double layer. {\it Phys. Rev. E} {\bf 1993}, {\it 48}, 3604. 
\bibitem {AttardJCP} McBride, A.; Kohonen, M.; Attard, P. The screening length of charge-asymmetric electrolytes: A hypernetted chain calculation. {\it J. Chem. Phys.} {\bf 1998}, {\it 109}, 2423-2428.
\bibitem {KjJCP2016} Kjellander, R. Nonlocal electrostatics in ionic liquids: The key to an understanding of the screening decay length and screened
interactions. {\it J. Chem. Phys.} {\bf 2016}, {\it 145}, 124503. 
\bibitem {Kj2020} Kjellander, R. A multiple decay-length extension of the Debye–H\"{u}ckel theory: to achieve high accuracy also for concentrated solutions and explain under-screening in dilute symmetric electrolytes. {\it Phys. Chem. Chem. Phys.} {\bf 2020}, {\it 22}, 23952-23985.
\bibitem {Kj1} Kjellander, R.; Mitchell, D. J. An exact but linear and Poisson-Boltzmann-like theory for electrolytes and colloid dispersions in the primitive model. {\it Chem. Phys. Lett.} {\bf 1992}, {\it 200}, 76-82. 
\bibitem {Kj2} Ulander, J.; Kjellander, R. The decay of pair correlation functions in ionic fluids: A dressed ion theory analysis of Monte Carlo simulations. {\it J. Chem. Phys} {\bf 2001}, {\it 114}, 4893-4904 
\bibitem {secmom} Stillinger, F. H.;  Lovett, R. Ion‐Pair Theory of Concentrated Electrolytes. I. Basic Concepts. {\it J. Chem. Phys.} {\bf 1968}, {\it 48}, 3858-3868.
\bibitem {Kirk1} Kirkwood, J. G. Statistical Mechanics of Liquid Solutions. {\it Chem. Rev.} {\bf1936}, {\it 19}, 275-307.
\bibitem {Kirk2} Kirkwood, J. G.; Poirier, J. C. The Statistical Mechanical Basis of the Debye–H\"{u}ckel Theory of Strong Electrolytes. {\it J. Phys. Chem.} {\bf 1954}, {\it 58}, 591-596.
\bibitem {PodWKB} Podgornik, R.; Zeks, B. Inhomogeneous Coulomb Fluid. A Functional Integral Approach. {\it J. Chem. Soc. Faraday Trans. 2} {\bf 1988}, \textit{84}, 611. 
\bibitem {NetzLoop} Netz, R. R.; Orland, H. Beyond Poisson-Boltzmann: Fluctuation effects and correlation functions. {\it Eur. Phys. J. E} \textbf{2000}, {\it 1}, 203-214.
\bibitem {NetzSC} Moreira, A.G.; Netz, R.R. Strong-coupling Theory for Counter-ion Distributions. {\it Europhys. Lett.} {\bf 2000}, \textit{52}, 705. 
\bibitem {NetzVir} Moreira, A.G.; Netz, R.R. Virial expansion for charged colloids and electrolytes. {\it Eur. Phys. J. D} {\bf 2002}, {\it 21}, 83–96. 
\bibitem {Podgornik2010} Kandu\v c, M.; Naji, A.; Forsman, J.; Podgornik, R. Dressed Counterions: Strong Electrostatic Coupling in the Presence of Salt. {\it J. Chem. Phys.} {\bf 2010} \textit{132}, 124701. 
\bibitem {Buyuk2020} Buyukdagli, S. Schwinger-Dyson Equations for Composite Electrolytes Governed by Mixed Electrostatic Coupling Strengths. {\it J. Chem. Phys.} {\bf 2020}, \textit{152}, 014902.  
\bibitem {netzvar} Netz, R. R.; Orland, H. Variational charge renormalization in charged systems. {\it Eur. Phys. J. E} {\bf 2003}, {\it 11}, 301-311. 
\bibitem {HatloVar} Hatlo, M. M.; Lue, L. A field theory for ions near charged surfaces valid from weak to strong couplings. {\it Soft Matter} {\bf 2009}, {\it 5}, 125-133. 
\bibitem {Demery} Demery, V.; Dean, D. S.; Podgornik, R. Electrostatic interactions mediated by polarizable counterions: Weak and strong coupling limits. {\it J. Chem. Phys.} {\bf 2012}, {\it 137}, 174903. 
\bibitem {Buyuk2023} Buyukdagli, S. Impact of the inner solute structure on the electrostatic mean-field and strong-coupling regimes of macromolecular interactions. {\it Phys. Rev. E} {\bf 2023}, {\it 107}, 064604.
\bibitem {DuncSolv} Coalson, R. D.; Duncan A.; Tal, N. B. Statistical Mechanics of a Multipolar Gas: A Lattice Field Theory. {\it J. Phys. Chem.} {\bf 1996}, {\it 100}, 2612-2620.
\bibitem {NL1} Buyukdagli, S.; Ala-Nissila, T.  Microscopic formulation of non-local electrostatics in polar liquids embedding polarizable ions. {\it Phys. Rev. E} {\bf 2013}, {\it 87}, 063201.
\bibitem {dipmem} Buyukdagli, S.; Podgornik. R. Contribution of dipolar bridging to phospholipid membrane interactions: A mean-field analysis.  {\it J. Chem. Phys.} {\bf 2021}, {\it 154}, 224902.
\bibitem {RudiPol} Podgornik, R. A variational approach to charged polymer chains: Polymer mediated interactions.  {\it J. Chem. Phys.} {\bf 1993}, {\it 99}, 7221-7231.
\bibitem {NetzPol} Netz, R. R. Strongly Stretched semi-flexible extensible polyelectrolytes and DNA. {\it Macromolecules} {\bf 2001}, {\it 34}, 7522–7529.
\bibitem {justin} Zinn-Justin, J. \textit{Quantum field theory and critical phenomena}; 2nd edition, Oxford University Press: 1993, Oxford.
\bibitem {JPCB2020} Buyukdagli, S. Nanofluidic Charge Transport under Strong Electrostatic Coupling Conditions. {\it J. Phys. Chem. B} {\bf 2020}, \textit{124}, 11299-11309.
\bibitem {Heyden2005} van der Heyden, D. J.; Stein, D.; Dekker, C. Streaming Currents in a Single Nanofluidic Channel. {\it Phys. Rev. Lett.} {\bf 2005}, {\it 95}, 116104.  
\bibitem {BuyukLang2022} Buyukdagli, S. Dielectric manipulation of polymer translocation dynamics in engineered membrane nanopores. {\it Langmuir} {\bf 2022}, \textit{38}, 122.  
\bibitem {BuyukDielDec} Buyukdagli, S. Explicit solvent theory of salt-induced dielectric decrement. {\it Phys. Chem. Chem. Phys.} {\bf 2022}, {\it 24}, 13976-13987.
\bibitem {math}  Abramowitz, M.; Stegun, I.A. {\it Handbook of Mathematical Functions}; Dover Publications: 1972, New York.
\bibitem {Exp} Hamer, W. J. ;  Wu, Y.-C. Osmotic Coefficients and Mean Activity Coefficients of Uni-univalent Electrolytes in Water at 25°C. {\it Journal of Physical and Chemical Reference Data} {\bf 1972}, {\it 1}, 1047-1099. 
\bibitem {rem1} The MC simulation data in Figs.~\ref{fig2}(b)-(d) and~\ref{fig4}(b)-(d) have been originally reported in Ref.~\cite{NetzMC} against the ionic packing fraction (the upper horizontal axis of our figures) and in term of the dimensionless ion sizes $d/\ell_{\rm B}$ (see the legends). In our plots, the latter have been transformed into the physical salt concentrations (the lower horizontal axis) and HC diameters at the liquid temperature and dielectric permittivity values indicated in the captions.
\bibitem {CS}  Carnahan, N. F.; Starling, K. E. Equation of State for Nonattracting Rigid Spheres. {\it J. Chem. Phys.} {\bf 1969}, {\it 51}, 635-636.
\bibitem {rem3} $\eta_{\rm c}$ has been defined for the osmotic coefficient in Fig.~\ref{fig3}(a) as well as the activity coefficient in Fig.~\ref{fig3}(b) as the packing fraction value where the deviation of the corresponding cumulant component (black curves) from its CS counterpart (red curve) exceeds $10\%$ of the latter.
\bibitem {book} Barthel, J.M.G.; Krienke, H.; Kunz, W. {\it Physical Chemistry of Electrolyte Solutions}; Springer Verlag: 1996, Darmstadt.
\bibitem{Buchner1999} Buchner, R.; Hefter, G. T.; May, P. M. Dielectric Relaxation of Aqueous NaCl Solutions. {\it J. Phys. Chem. A} {\bf 1999}, {\it 103},  1-9. 
\bibitem{Sridhar1990} Wei, Y.-Z.; Sridhar, H. Dielectric spectroscopy up to 20 GHz of LiCl/${\rm H}_2$O solutions.  {\it J. Chem. Phys.} {\bf 1990}, {\it 92}, 923-926. 
\bibitem{Wei1992} Wei, Y.-Z.; Chiang, P.; Sridhar, S. Ion size effects on the dynamic and static dielectric properties of aqueous alkali solutions. {\it J. Chem. Phys.} {\bf 1992}, {\it 96}, 4569-4573. 
\bibitem {Gavish} Gavish, N.; Promislow, K. Dependence of the dielectric constant of electrolyte solutions on ionic concentration: A microfield approach.  {\it Phys. Rev. E} {\bf 2016}, {\it 94}, 012611.
\bibitem {rem2} The experimental molal ion activities $\gamma_m$ in Ref.~\cite{Exp} have been provided in terms of the molality $m$ of the solution. In Fig.~\ref{fig5}, the molal activities have been converted to molar activities $\gamma$ via the identity $\gamma/\gamma_m=10^{-3}\rho_{\rm w}m/n_i$~\cite{Exp}, and the molality values have been transformed into molarities $ n_i$ with the use of the relation
\be
\label{r1}
 n_i=10^{-3}\frac{m\rho_{\rm el}( n_i)}{1+mm_{\rm s}}.
\ee
In the formulas above, $\rho_{\rm w}$ and $\rho_{\rm el}( n_i)$ are respectively the mass density of pure water and the electrolyte taken from Ref.~\cite{ElDen}, and $m_{\rm s}$ stands for the molar mass of the corresponding salt. All parameters are in SI units. 
\bibitem {Hof1} Hofmeister, F. Zur Lehre von der Wirkung der Salze. {\it Arch. Exp. Pathol. Pharmakol.} {\bf 1888}, {\it 24}, 247. 
\bibitem {Hof2} Cacace, M. G.; Landau, E. M.; Ramsden, J. J. The Hofmeister series: salt and solvent effects on interfacial phenomena. {\it Quarterly Reviews of Biophysics} {\bf 1997}, {\it 30}, 241-277.
\bibitem {ElDen} Novotny, P.; Sohnel, O. Densities of binary aqueous solutions of 306 inorganic substances. {\it J. Chem. Eng. Data} {\bf 1988}, {\it 33}, 49-55.
\end{thebibliography}
\end{document}